\documentclass[prx,aps,twocolumn,amsmath,amssymb]{revtex4}

\pdfoutput=1
\usepackage{hyperref}
\usepackage{graphicx}
\usepackage[usenames]{color}
\usepackage{bm}

\def\cal#1{\mathcal{#1}}
\def\eqq#1{Eq.~(\ref{#1})}
\def\eq#1{(\ref{#1})}
\def\av#1{\langle #1 \rangle}

\def\f#1{Fig.~\ref{#1}}

\def\c#1{~\cite{#1}}
\def\cc#1{Ref.\c{#1}}

\def\pp{P\'{e}clet~}

\def\s#1{Section~\ref{#1}}
\def\kt{k_{\rm B}T}

\def\beq{\begin{equation}}
\def\eeq{\end{equation}}
\def\bea{\begin{eqnarray}}
\def\eea{\end{eqnarray}}

\begin{document}

\title{On the microscopic origin and macroscopic implications of\\ lane formation in mixtures of oppositely-driven particles}
\author{Katherine Klymko$^1$}
\author{Phillip L. Geissler$^{1,2}$}
\email[]{geissler@berkeley.edu}
\author{Stephen Whitelam$^3$}
\email[]{swhitelam@lbl.gov}
\affiliation{$^1$Department of Chemistry, University of California, Berkeley, CA 94720, USA \\$^2$Chemical Sciences Division, Lawrence Berkeley National Laboratory, Berkeley, CA 94720, USA \\$^3$Molecular Foundry, Lawrence Berkeley National Laboratory, 1 Cyclotron Road, Berkeley, CA 94720, USA}

\begin{abstract}
Colloidal particles of two types, driven in opposite directions, can segregate into lanes [Vissers et al. Soft Matter 7, 2352 (2011)]. This phenomenon can be reproduced by two-dimensional Brownian dynamics simulations of model particles [Dzubiella et al. Phys. Rev. E 65, 021402 (2002)]. Here we use computer simulation to assess the generality of lane formation with respect to variation of particle type and dynamical protocol. We find that laning results from rectification of diffusion on the scale of a particle diameter: oppositely-driven particles must, in the time taken to encounter each other in the direction of the drive, diffuse in the perpendicular direction by about one particle diameter. This geometric constraint implies that the diffusion constant of a particle, in the presence of those of the opposite type, grows approximately linearly with \pp number, a prediction confirmed by our numerics over a range of model parameters. Such environment-dependent diffusion is statistically similar to an effective interparticle attraction; consistent with this observation, we find that oppositely-driven non-attractive colloids display features characteristic of the simplest model system possessing both interparticle attractions and persistent motion, the driven Ising lattice gas [Katz, Leibowitz, Spohn, J. Stat. Phys. 34, 497 (1984)]. These features include long-ranged correlations in the disordered regime, and a critical regime characterized by a change in slope of the particle current with \pp number and by fluctuations that grow with system size. By analogy, we suggest that lane formation in the driven colloid system is in the macroscopic limit a phase transition, but that macroscopic phase separation would not occur in finite time upon starting from disordered initial conditions. 
\end{abstract}

\maketitle

\section{Introduction} 
\label{intro}

Systems driven out of equilibrium display a rich variety of patterns\c{domb1995statistical,ramaswamy2010mechanics}. Here we study patterns formed by a two-dimensional, two-component colloidal mixture of overdamped particles in which one species (`red') possesses a bias to move persistently in one direction, and the other species (`blue') possesses a bias to move persistently in the opposite direction. L\"{o}wen and coworkers have shown that for large enough bias such particles form persistently-moving lanes, extended in the direction of the bias, segregated by particle type\c{dzubiella2002lane,chakrabarti2003dynamical,chakrabarti2004reentrance,kohl2012microscopic}. Lane formation is seen in three-dimensional experiments of binary colloidal mixtures driven by an electric field\c{vissers2011lane}, and in driven binary plasmas\c{ivlev2012complex, sutterlin2010lane}. Much is already known about the microscopic origin of laning in model systems and its macroscopic manifestation. On the microscopic side Chakrabarti et al. used  dynamic density functional theory to argue that Langevin dynamics of oppositely-driven particles implies laning via a dynamic instability of the homogenous phase\c{chakrabarti2003dynamical,chakrabarti2004reentrance}; Kohl et al. showed, using a many-body Smoluchowski equation for interacting Brownian particles, that driven systems in the homogeneous phase display anisotropic pair correlations that foreshadow the onset of laning\c{kohl2012microscopic}. On the macroscopic side Glanz et al. used large-scale numerical simulations to show that characteristic lengthscales in the model grow (at large drive speed) exponentially or algebraically with drive speed\c{glanz2012nature}. The authors of that work suggested that lane formation in two dimensions is therefore not a true phase transition.

In order to assess the generality of lane formation, i.e. to determine if laning persists upon changing the type of particle and the dynamical rules used, we modeled oppositely-driven particles using three distinct numerical protocols. The first (Protocol I) comprises soft particles in continuous space evolved by Langevin dynamics, similar to protocols used by other authors\c{dzubiella2002lane}. The second (Protocol II) comprises hard particles in continuous space evolved by Monte Carlo dynamics. The third (Protocol III) comprises lattice-based particles evolved by Monte Carlo dynamics. Isolated particles under all protocols move diffusively and possess a positive drift velocity $V$ to the left or to the right of the simulation box. Left-movers (red particles) and right-movers (blue particles) are equally numerous. 

We used Protocol I to reproduce the basic phenomenology of laning studied by other authors: for large enough $V$ (or, equivalently, \pp number), persistently-moving red and blue lanes form. Protocol II can reproduce this phenomenology, but only if the basic step size of the Monte Carlo protocol is a small fraction of the particle diameter; otherwise, the protocol results in jammed bands that point perpendicular to the direction of biased motion. Under Protocol III, upon increase of \pp number, only jamming occurs.

From comparison of these protocols we draw three conclusions. The first relates to the microscopic origin of laning: because it occurs for soft and hard particles, and under distinct dynamic protocols, laning can be considered to be a statistical effect that results from the following simple geometric constraint. In order not to overlap, oppositely-driven particles must, in the time taken to meet each other in the direction of the drive, diffuse laterally (perpendicular to the drive) by about a particle diameter. In other words, diffusion on the scale of a particle diameter is rectified or ratcheted in the manner demonstrated in \f{fig0}. Particles then possess a lateral diffusion constant that scales linearly with drift speed $V$ (or, equivalently, \pp number) at large $V$, and approximately as the square root of the local density of particles of the opposite type. This diffusion constant can exceed that of a particle surrounded by particles of the same type, implying a tendency to form lanes. Enhanced diffusion of particles in the presence of oppositely-moving particles was identified to be the origin of laning in the simulations and experiments of \cc{vissers2011lane}, and similar mechanisms have been described for pattern formation in systems of agitated particles\c{grunwald2016exploiting}. Our first conclusion complements this work by identifying the geometric origin of the phenomenon and revealing the scaling of diffusion enhancement with \pp number.

Our second conclusion relates to the macroscopic consequences of laning, and follows from the first conclusion via a connection between environment-dependent diffusion rates and effective interparticle attractions. Lane formation results from the fact that particles possess environment-dependent diffusion rates. One can show that a set of hard particles that possess environment-dependent diffusion rates is equivalent to a set of attractive particles (see e.g.\ Ref.\c{sear2015out}) whose interaction energies scale logarithmically with diffusion rates. One can therefore consider the driven model to possess both persistent motion and effective interparticle attractions. The simplest model system possessing both features is the driven Ising lattice gas (DLG), also known as the Katz-Lebowitz-Spohn model\c{katz1984nonequilibrium,zia2010twenty}. We show here that the two models have strong qualitative similarities. The DLG displays long-ranged correlations in the disordered phase; we show numerically that the same is true of the off-lattice model. The DLG also displays a continuous order-disorder phase transition (in a non-Ising universality class) between a disordered phase and a phase characterized by lane-like structures\c{schmittmann1990critical,levine2001ordering,saracco2003critical,daquila2012nonequilibrium}. This transition is characterized by a break in the slope of particle current with model parameters, and system-spanning fluctuations. We show that the same is true of the off-lattice model. 

\begin{figure}[]
	\includegraphics[width=0.9\linewidth]{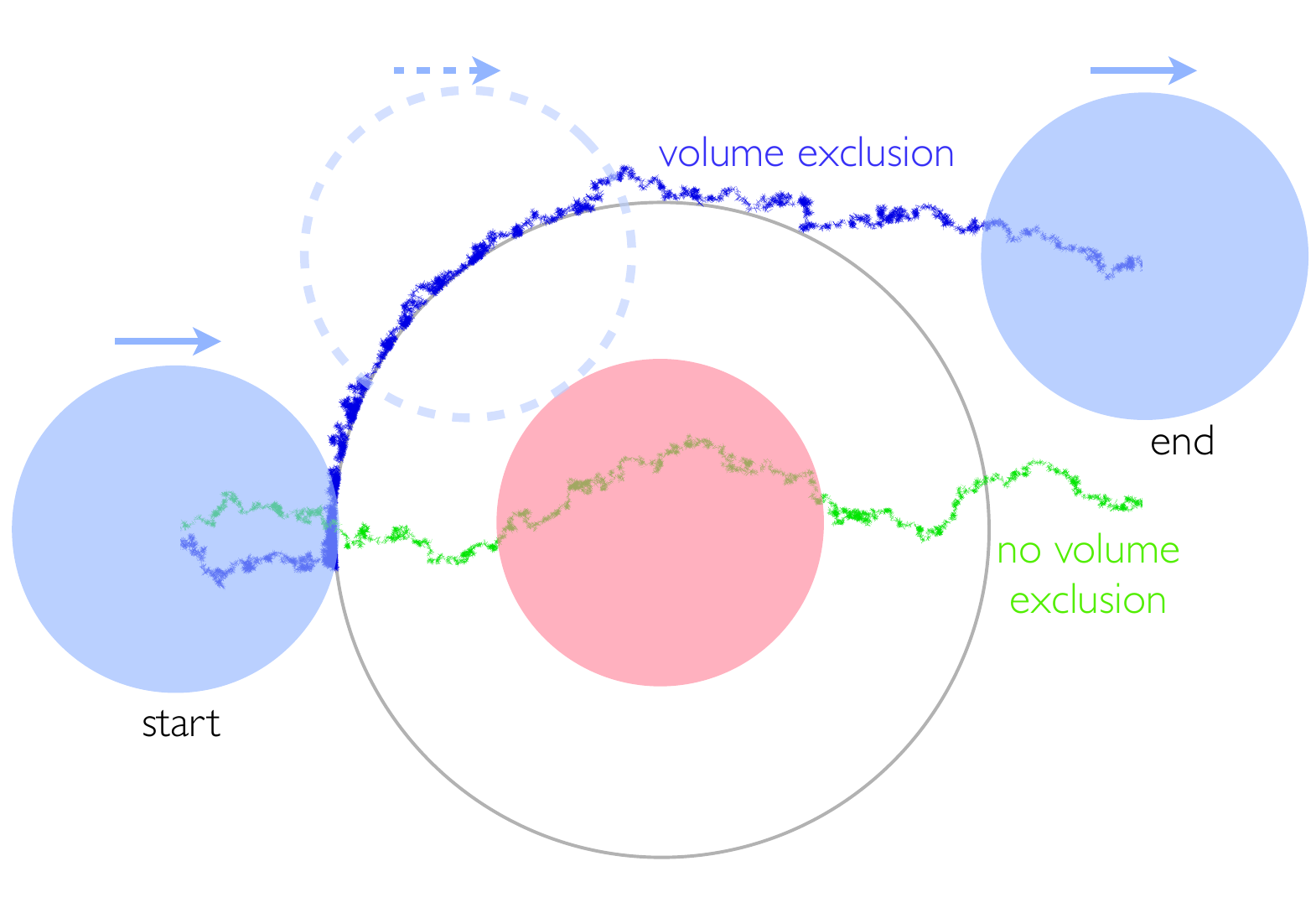}
	\caption{\label{fig0} Rectification of diffusion on the scale of a particle diameter underpins the laning transition. Red and blue particles moving persistently in opposite directions (left-right) must diffuse laterally (up-down) by about a particle diameter in the time taken to encounter each other. Driven rightward diffusion of the blue particle in the rest frame of the red particle (at center) shows the rectification of lateral diffusion that occurs if red and blue particles exclude volume (blue line); the large gray circle indicates the position of closest possible approach of the centers of red and blue particles. The green trajectory shows similar driven diffusive motion in the absence of volume exclusion. Both trajectories were generated using the dynamics described in Section S1.1.}
\end{figure}

Continuing this analogy to its conclusion, we expect lane formation in a macroscopic version of the off-lattice driven system to be a true phase transition. Although this conjecture appears to contradict the conclusion of Ref.\c{glanz2012nature}, that laning should emerge only as a smooth crossover in the thermodynamic limit, we believe that the two statements are consistent. The simulations of Ref.\c{glanz2012nature} used disordered initial conditions, and it has been shown that the time taken for the DLG to relax to its steady state diverges with system size upon starting from disordered initial conditions\c{levine2001ordering}. The analogy we have drawn therefore suggests that macroscopic domains in the off-lattice model would persist if built `by hand' (provided that the aspect ratio of the system is chosen `correctly', see e.g. \cc{zia2000possible}), but would indeed not be seen in finite time upon starting from disordered initial conditions, consistent with the conclusion of Ref.\c{glanz2012nature}. We present numerical evidence to support this conjecture. Considering that the off-lattice model\c{dzubiella2002lane} can reproduce the basic phenomenology of lane formation seen in experiments\c{vissers2011lane}, the comparison we have drawn between the off-lattice system and the DLG suggests that the latter may have application to experiment (indeed, previous studies of related models were done with ionic conductors in mind\c{murch1977computer}).

Our third conclusion relates to numerical modeling of driven systems: the qualitative outcome of our driven simulations appears to be more sensitive to protocol than is the simulation of undriven systems. It is well known that Monte Carlo dynamics of undriven particles, in the limit of zero step size, is formally equivalent to a Langevin dynamics\c{tiana2007use,kikuchi1991metropolis}. As suggested by this equivalence, undriven systems evolved under Monte Carlo dynamics with finite step size often behave qualitatively like their Langevin-evolved counterparts\c{tiana2007use,sanz2010dynamic}, even if not identical in all aspects of their dynamics\c{whitelam2007avoiding}. In the present study the same is true only if the basic step size of the Monte Carlo procedure (Protocol II) is extremely small. As step size increases, the tendency to lane is less strong -- laning results from enhancement of diffusion on scales less than a particle diameter, and such motion is less accurately represented as step size increases -- and the tendency to jam is stronger. Monte Carlo protocols carried out with step size above a certain value therefore show no laning at all, in contrast to Langevin simulations. For the lattice-based Protocol III, the tendency to lane is entirely absent, because the basic step size is equal to that of the particle diameter. Our results therefore highlight the subtleties of modeling driven systems using different protocols.

 In our model, the rectified flow of colloids results in pattern formation. Similar flow effects result in effective attractions in other driven models\c{penna2003dynamic,dzubiella2003depletion} and can lead to the onset of lane-like patterns\c{brader2011density}. 

In \s{model} we introduce the numerical protocols that we have studied. In \s{results} we compare their behavior, and there and in \s{ising} we describe how this comparison implies the conclusions stated above. We summarize our results in \s{conc}.

\section{Numerical models of oppositely-driven particles}
\label{model}

We considered three numerical protocols, two off-lattice (Protocols I and II) and one on-lattice (Protocol III). In Protocol I particles were evolved using Langevin dynamics, while in Protocols II and III particles were evolved using Monte Carlo dynamics. In all protocols we considered two types of particle, labeled red and blue, which are confined to two spatial dimensions and which interact repulsively. Particles undergo diffusion biased such that red particles possess a drift to the left, and blue particles possess an equal drift to the right. Our simulation boxes (generally) were periodic in both directions, and we focused on patterns generated using equal numbers of red and blue particles. We considered systems over a range of densities and \pp numbers (Pe). Density is defined off-lattice as $\rho=\frac{N}{A}$, where $N$ is the total number of particles and $A$ is the system area, and on-lattice as the fraction of occupied lattice sites. \pp number is defined for Protocol I as the ratio of the magnitude of biasing force to the thermal energy, $F_{\rm ex}\sigma/(\kt)$, where $\sigma$ is the particle diameter. \pp number is defined for Protocols II and III as the combination $v_{x}\sigma/D_0$ of the (bare) particle drift velocity, diffusion constant, and particle diameter. All distances are given in units of $\sigma$. Our protocols do not take into account hydrodynamic interactions\c{rex2008influence} which may have important effects in experimental realizations of this system.

{\em Protocol I: Langevin dynamics.} The state of the system is represented by the positions of all the particles $\{ \textbf{r}_i \}$. Particles are disks with diameter $\sigma$. Each particle undergoes overdamped Langevin dynamics governed by 
\beq
\label{lang_eq}
\dot{\bold{r}_i}=D\beta \left[\bold{F}_{ \rm rep} \left( \{ \bold{r}_i \} \right) + \bold{F}_{\rm ex} \right]+\sqrt{2D}\bm{ \eta }_i(t).
\eeq
Here $\bold{F}_{ \rm rep}$ is an excluded-volume repulsive force derived from the WCA potential, which reads
\beq
V(r_{ij})=4\epsilon \left[ \left(\frac{\sigma}{r_{ij}}\right)^{12}-\left(\frac{\sigma}{r_{ij}}\right)^6 + \frac{1}{4}\right]
\eeq
if $r_{ij}<2^{1/6}\sigma$, and zero otherwise. We take $\epsilon$ as our unit of energy. $\bold{F}_{ex}$ is a constant force acting only in the $x$ direction. For red particles this force is $\beta \bold{F}_{ex}=-\frac{\text{Pe}}{\sigma} \hat{e}_x$ and for blue particles $\beta \bold{F}_{ex}=\frac{\text{Pe}}{\sigma} \hat{e}_x$ where Pe denotes the \pp number. $D$ in \eqq{lang_eq} is the bare translational diffusion constant (which we refer to as the bare diffusion constant in the text), $\beta\equiv \frac{1}{\kt}$, and the $\bm{ \eta }_i=(\eta_i^x,\eta_i^y)$ are white noise variables with $\langle \bm{\eta}_i(t) \rangle =0$ and $\langle \bold{\eta}^{\mu}_i(t) \eta^{\nu}_j(t') \rangle=\delta_{ij}\delta_{\mu \nu}\delta(t-t')$. Simulations had a maximum timestep of $10^{-5} \Delta t$ where $\Delta t \equiv \sigma^2/D$ was our unit of time. We used LAMMPS\c{plimpton1995lammps} to integrate the equations of motion. 

{\em Protocol II: Monte Carlo dynamics (off-lattice).} Monte Carlo (MC) simulations off-lattice employed single-particle Metropolis moves with particle displacements chosen to effect a drift of red and blue particles in opposite directions. We determined the connection between displacement parameters and an isolated particle's \pp number and diffusion constant as described in \s{monte}. The resulting mapping depends on the basic displacement scale (step size). We ran simulations for hard disks and for WCA pair particles, for a range of step sizes.

{\em Protocol III: Monte Carlo (on-lattice).}  We considered volume-excluding particles present at a range of densities on a square lattice. The dynamics, which conserve particle number, consisted of choosing a particle at random and moving it to one of the four neighboring sites with biased probability in the driven direction and equal probability in the lateral directions (see section S1 for more details). Moves that take particles to already occupied sites were rejected. This lattice model was originally studied in\c{schmittmann1992onset}.
\begin{figure*}[]
	\includegraphics[width=0.8\linewidth]{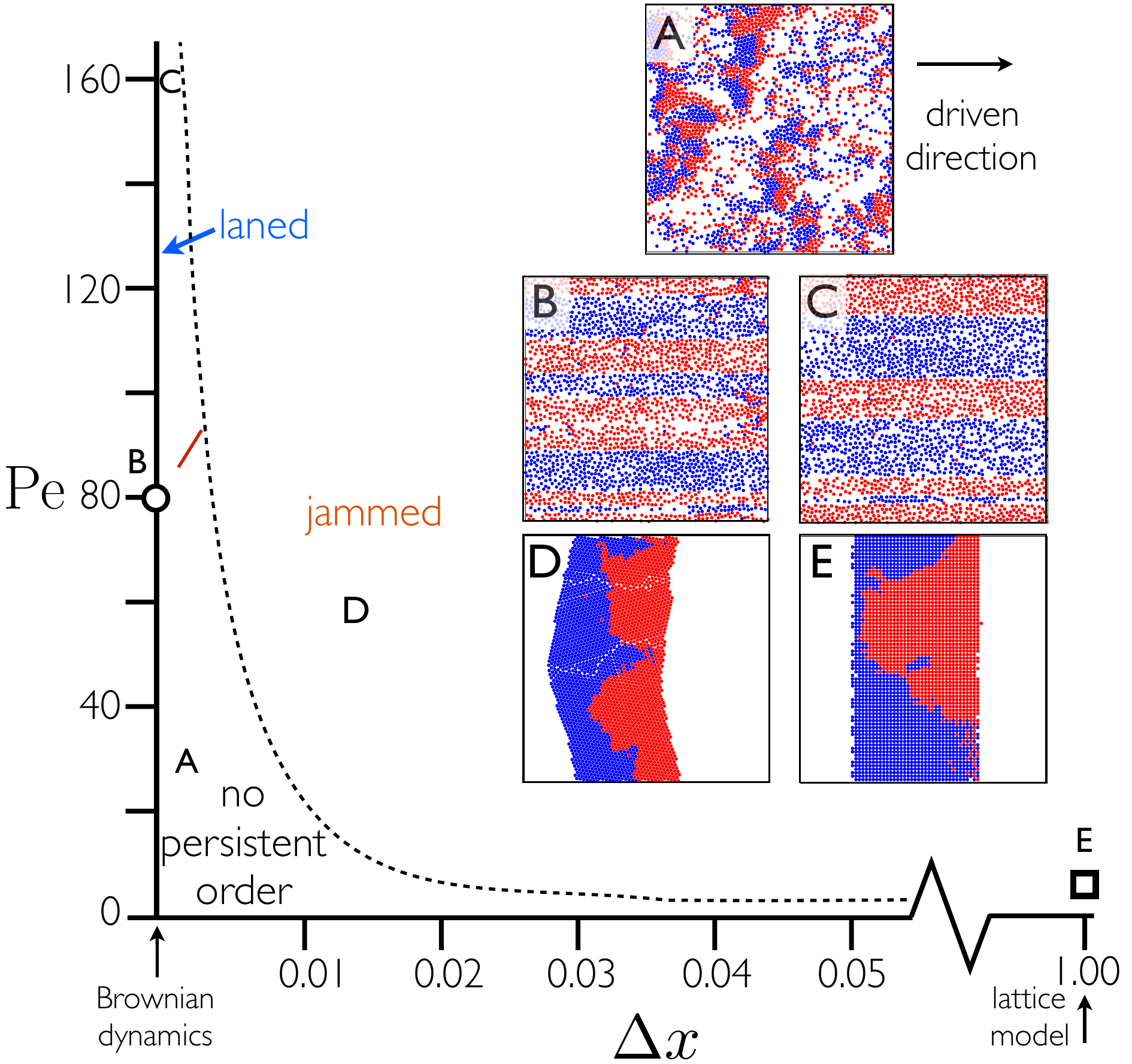}
	\caption{\label{fig1} Steady-state dynamic regimes observed using our three numerical protocols for systems with 2500 particles, starting from disordered initial conditions at density $\rho = 0.5$. The vertical axis is \pp number. The horizontal axis interpolates between the Langevin simulations of Protocol I (at `zero step size', i.e. $\Delta_x=0$) and the lattice-based simulations of Protocol III ($\Delta_x=1$), with the results of off-lattice hard-disk Monte Carlo simulations (Protocol II) shown for a range of intermediate step sizes $\Delta_x$. Snapshots A, C,  and D were obtained using Protocol II; B was obtained using Protocol I; and E was obtained using Protocol III. In the snapshots, the driven direction is left-right. The lines show the approximate boundaries between different steady state behaviors: the black dashed line shows the boundary between the jammed and flowing states and the solid red line shows the boundary between disordered and laned states (see section S1 for more details). Monte Carlo simulations reproduce the results of Langevin simulations if the basic step size of the former is small enough; otherwise, Monte Carlo and Langevin results differ qualitatively. Lattice-based Monte Carlo simulations jam for \pp number of order unity.}
\end{figure*}

\begin{figure}[]
	\includegraphics[width=\linewidth]{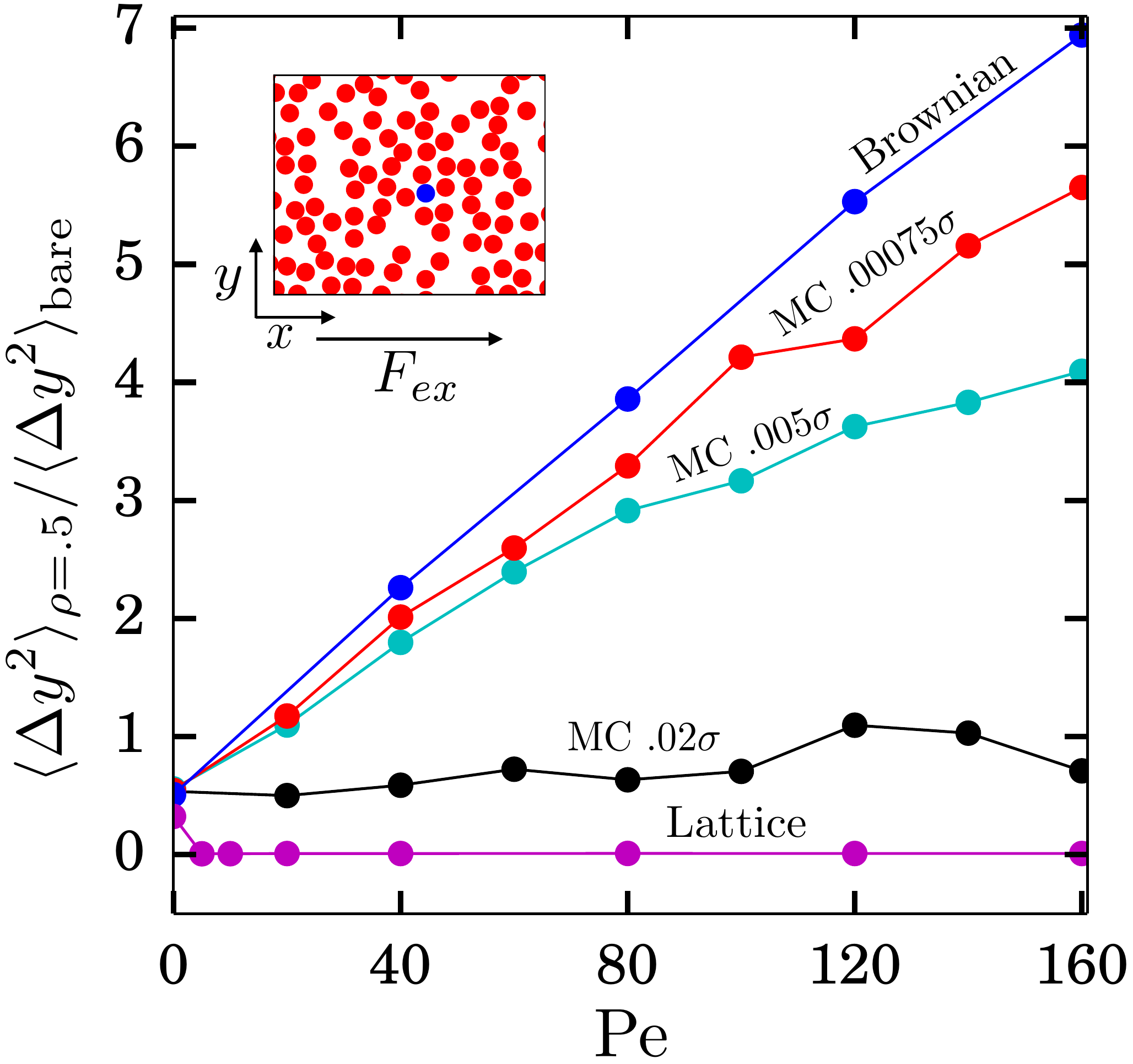}
	\caption{\label{fig2} The lateral diffusion constant (perpendicular to the driven direction) as a function of Pe for a test particle in the presence of particles of the other type. We have normalized diffusion constant by the bare value. The longitudinal component of diffusion behaves similarly; see \f{figSI2}. In these simulations one blue particle is placed in a box of red particles, and the blue particle's diffusion constant is measured. This enhancement of diffusion with Pe of one particle in the presence of particles of the other type underpins the laning transition, seen for Langevin simulations and off-lattice Monte Carlo simulations with sufficiently small basic step size. Off-lattice Monte Carlo protocols with larger step sizes do show diffusion enhancement with Pe, but tend to jam rather than to exhibit laning. On-lattice Monte Carlo protocols show no enhancement of diffusion constant with \pp number, and jam readily.}
\end{figure}

\begin{figure}[]
	\includegraphics[width=\linewidth]{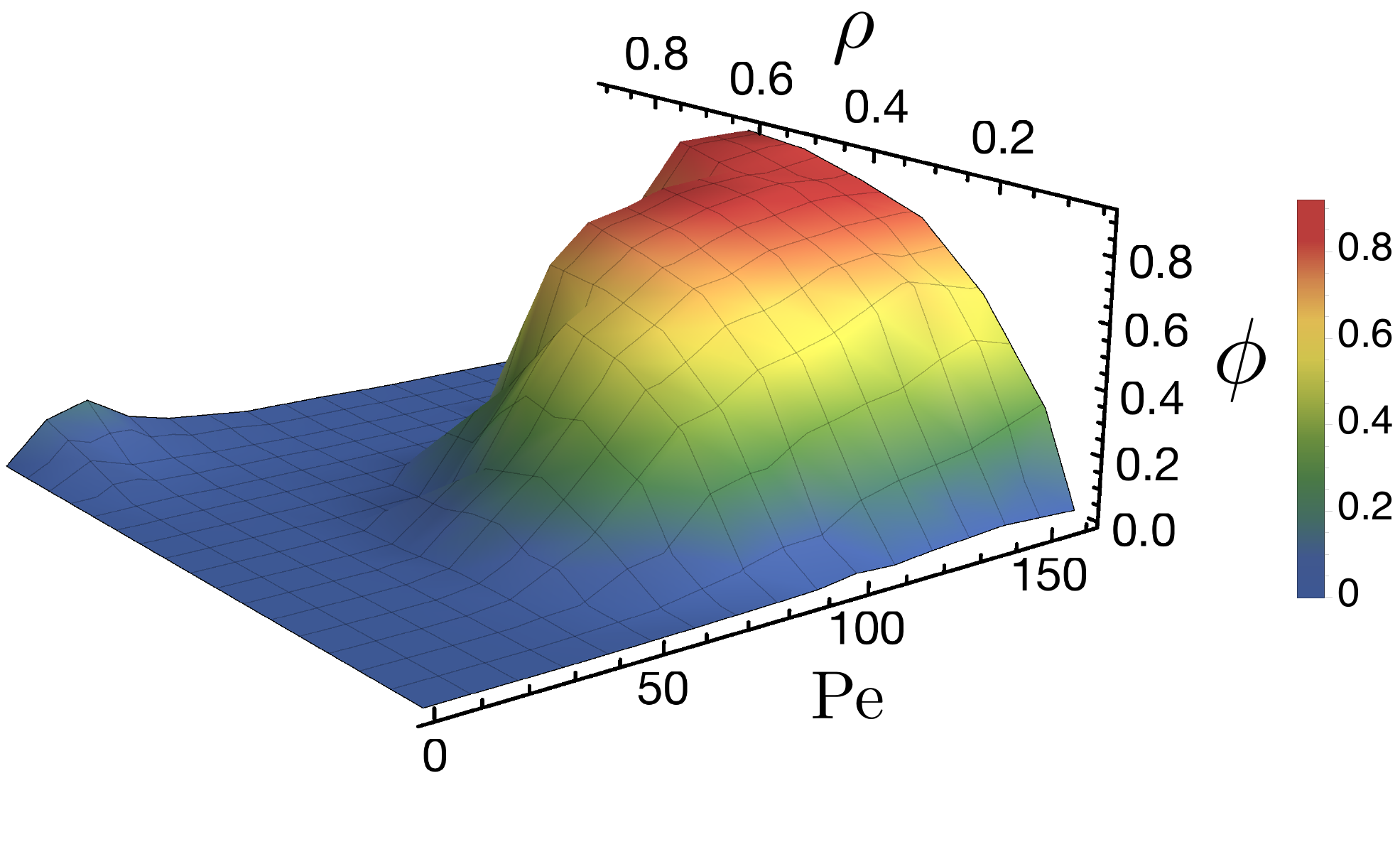}
	\includegraphics[width=\linewidth]{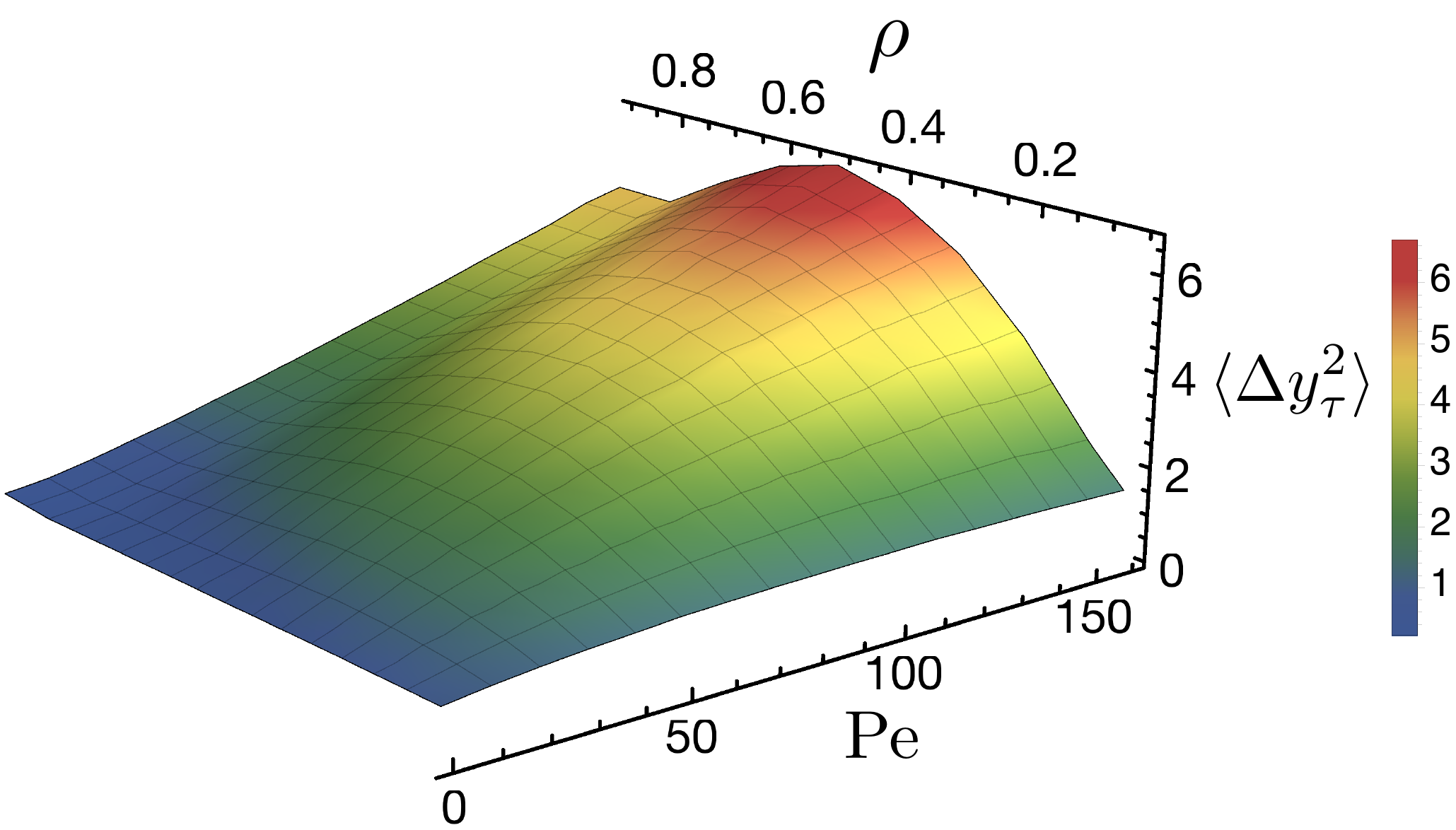}
	\caption{\label{fig3} The enhancement of lateral diffusion as a function of \pp number for one particle in the presence of particles of the opposite type (bottom panel) correlates approximately with an increase in an order parameter $\phi$ for lane formation in equimolar mixtures (top panel). Both calculations used Langevin dynamics (Protocol I). The bottom panel comes from the steady-state simulations of \f{fig2}, in which a test particle is placed in a simulation box containing only particles of the other type. The top panel comes from the equimolar red-blue mixture simulations of \f{fig1}.}
\end{figure}

{\em Order parameters.} We characterized the dynamics and structures within simulations using the averaged particle activity
\beq
A(\tau)\equiv \left\langle \frac{1}{\tau N_{\rm tot}} \sum_{i=1}^{N_{\rm tot}}  |x_i(t+\tau)-x_i(t)| \right\rangle,
\eeq
 where $N_{\rm tot}$ is the total number of particles. We also used the structural order parameter
\beq
\phi \equiv \left \langle \frac{1}{N_{\rm red}} \sum_{i=1}^{N_{\rm red}}\prod_{j=1}^{N_{\rm blue}} \theta \left( |y_i-y_j| -\frac{ {\rho}^{-1/2} }{2} \right) \right\rangle
\eeq
used by other authors to characterize laning\c{dzubiella2002lane,glanz2012nature}. $\phi$ in effect counts the percentage of particles in a lane-like environment. The brackets for both order parameters indicate a time average. 

Systems were considered to be `jammed' if the average activity at steady state dropped below half that of an isolated particle. Systems were considered to be laned if (a) the average activity was greater than half that of an isolated particle and  (b) $\phi$ was greater than a particular value, usually 0.5 (see \f{figSI1} for plots of these order parameters as a function of time).

\section{Comparison of numerical protocols}
\label{results}

\begin{figure*}[]
	\includegraphics[width=\linewidth]{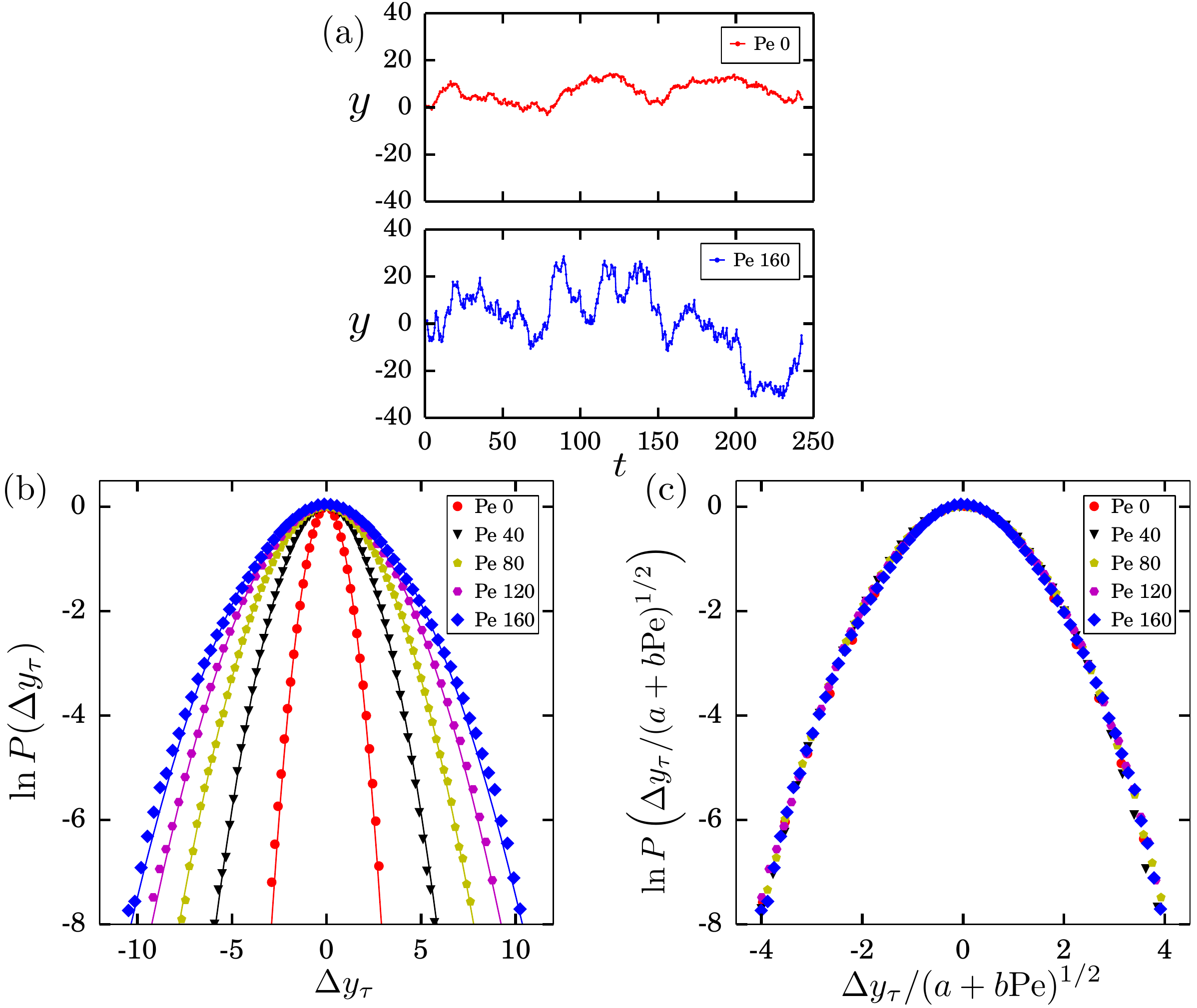}
	\caption{\label{fig4} (a) Trajectories of the $y$-coordinate of a single blue particle placed in a periodic box of red particles present at density $\rho=0.5$ (see \f{fig2}), for ${\rm Pe}= 0$ and ${\rm Pe}= 160$. Particles are driven in the $x$-direction. These trajectories show visually the enhanced diffusion in the presence of the drive. (b) Histograms of lateral diffusion constant for various Pe can be collapsed (c) by rescaling $\Delta y_{\tau}$ by $(a + b \, {\rm Pe})^{-1/2}$, where $a$ and $b$ are constants, as suggested by \eqq{diff_const}. This collapse indicates that the simple physical argument that gives rise to \eqq{diff_const} captures, in this parameter regime, the microscopic origin of enhanced diffusion.}
\end{figure*}

{\bf Numerical protocols show a range of qualitative behavior.} In \f{fig1} we identify the steady-state dynamic regimes obtained using our three dynamic protocols in the space of \pp number versus protocol type. The limit of zero step size, $\Delta_x=0$ on the horizontal axis, corresponds to Langevin dynamics simulations, whose results are similar to those published by other authors\c{dzubiella2002lane,chakrabarti2003dynamical,chakrabarti2004reentrance,vissers2011lane,kohl2012microscopic}: we observe a transition from a disordered mixture to persistently-moving lanes of like-colored particles parallel to the driven direction at a \pp number of about 80. We shall refer to the value of \pp number at the transition as the critical \pp number. Off-lattice Monte Carlo simulations with sufficiently small step size show qualitatively similar behavior. For small step size the critical \pp number seen in these simulations is similar to the Langevin value. As step size is increased the Monte Carlo critical \pp number increases, and the laning transition eventually disappears: simulations run using a basic step size above some threshold show qualitatively different behavior to Langevin simulations, forming `jammed' stripes perpendicular to the direction of the external field~\footnote{The tendency to jam results from the reduction of a particle's drift velocity with density; this reduction rate increases with Monte Carlo step size, and when large enough can trigger density-density phase separation\c{tailleur2008statistical}.}. This threshold corresponds to a basic displacement of 1\% of a particle diameter or less ($10^{-3}\sigma$ for hard disks and $10^{-2}\sigma$ for WCA particles), which is rather small for Monte Carlo simulations: for {\em undriven} systems one can sometimes obtain approximate dynamical realism using Monte Carlo simulations with much larger basic step size\c{berthier2007revisiting}. It is notable, given recent interest in modeling driven and active systems, that small changes in dynamic protocol can change the apparent steady state of a system of driven particles. On-lattice Monte Carlo simulations ($\Delta_x = 1$) also formed jammed perpendicular stripes as Pe is increased, rather than lanes parallel to the direction of driving.\\

{\bf Laning results from enhanced diffusion of particles in the presence of particles of the other type.} \f{fig1} shows that Langevin simulations of soft particles and Monte Carlo simulations of hard (and soft) particles, for small enough step size, exhibit similar phenomenology. Such similarities indicate that the origin of laning can be understood without reference to fine details of the system under study. A detail-insensitive mechanism for lane formation is suggested by Ref.\c{vissers2011lane}, which showed that particles undergoing lane formation experienced time-dependent diffusion constants that are large when the system is disordered, and become smaller when the system forms lanes. In order to understand how particle mobilities are affected by a driven environment in a more controlled setting we measured diffusion rates of particles at steady state, by measuring the diffusion constant of a blue `test' particle placed in a periodic simulation box in which only red particles are present. Such pseudo-single-particle simulations allowed us to isolate the effects of the drive without the complication of attendant pattern formation. In \f{fig2} we show the lateral component of the blue particle's diffusion constant for our three numerical protocols. An enhancement of diffusion constant with \pp number is seen in all cases except for the on-lattice Monte Carlo simulations. In \f{fig3} we show for Langevin simulations that this enhancement of diffusion, measured in a steady-state, quasi-single-particle simulation, correlates approximately with the onset of laning measured in an equimolar mixture of red and blue particles.\\

{\bf Enhanced diffusion follows from simple geometric constraints.} \f{fig2} demonstrates that enhanced diffusion of particles in the presence of those of the opposite type occurs for different interaction potentials and dynamic schemes. Such robustness suggests a simple geometric origin for the effect, summarized graphically in \f{fig0}, which we quantify in the following way. In order to avoid overlapping, two oppositely-colored particles must diffuse laterally by about one particle diameter in the time taken for them to encounter each other in the direction of drift. Such avoidance implies an enhancement of a particle's diffusion constant. To see this, consider the equation of motion of the $x$-coordinate of a particle undergoing driven Brownian motion,
\beq
\dot{x}(t) = V + \sqrt{2D} \eta(t).
\eeq
Here $V$ and $D$ are the drift velocity and diffusion constant of the particle, and $\eta$ is a Gaussian white noise with zero mean and unit variance. For a particle initially at the origin we have
\beq
\label{eq1}
\av{x(t)^2} = (Vt)^2+ 2D t,
\eeq
where $\av{\cdot}$ denotes an average over noise. Let the characteristic distance in the driven direction between the center of the test particle and one of the opposite color be $\ell$ (we expect roughly $\ell^{-1} \propto \sqrt{\rho(1-\chi)}$, where $\rho$ is the mean number of particles per unit area and $\chi$ is the fraction of particles in the test particle's neighborhood of its own type). The characteristic encounter time $\tau$ of the two particles can be found from \eq{eq1} by setting $\av{x(\tau)^2} = (\ell/2)^2$, giving
\beq
\label{eq2}
V^2 \tau^2+ 2D \tau - (\ell/2)^2=0.
\eeq
If in time $\tau$ we require our test particle to diffuse laterally by a distance of order one particle radius, $\sigma/2$, so as to avoid overlap, then it must have an effective lateral diffusion constant of order $D_{\rm eff}(V) = \sigma^2/(8\tau)$, i.e.
\beq
\label{eq3}
D_{\rm eff}(V) = \frac{\sigma^2}{8} \frac{V^2}{\sqrt{D^2 + \ell^2 V^2/4}-D},
\eeq
upon solving \eq{eq2} for $\tau$.

For large $V$ we have
\beq
D_{\rm eff}(V) \approx  \frac{\sigma^2}{4 \ell} V +\frac{\sigma^2}{2 \ell^2}D.
\eeq
Assuming that the drift speed $V$ of the particle is equal to its bare drift velocity $V_0$ (which our numerics indicates is approximately true under conditions for which lanes form), we have $V = V_0 \equiv  D_0 {\rm Pe}/\sigma$ and
\beq
\label{diff_const}
D_{\rm eff}({\rm Pe}) \approx \frac{\sigma}{4 \ell} D_0 {\rm Pe} + \frac{\sigma^2}{2 \ell^2}D.
\eeq
Thus we predict that rectification of diffusion in the presence of particles of the opposite type results in an effective diffusion constant that increases, at large \pp number, linearly with \pp number (here we assume that $D$ does not vary with Pe). In \f{fig4} and \f{figSI4} we show that the linear dependence of diffusion constant with Pe predicted by \eqq{diff_const} is indeed seen in our steady-state simulations across a range of model parameters. In physical terms \eqq{diff_const} indicates that particles experience a net flux that takes them from a domain of oppositely-colored particles to a domain of like-colored ones. Such a flux implies a basic tendency for formation of domains of persistently-moving like-colored particles, i.e. lanes, although this equation does not indicate for which Pe this will happen. 

For weak driving (small $V \approx V_0$) we expect linear scaling to break down; there, we can expand \eq{eq3} to get
\beq
\label{small_pe}
D_{\rm eff}({\rm Pe}) \approx \frac{\sigma^2}{\ell^2}D + \frac{1 }{16 }D_0 ({\rm Pe})^2,
\eeq 
suggesting that for small Pe the effective diffusion constant of a particle in the presence of those of the opposite type increases quadratically with Pe (we might expect the observed diffusion constant of a particle to be the larger of \eq{small_pe}, and $D$). Such breakdown of linear scaling at weak driving is consistent with our simulations: see \f{figSI4}.
 
Note that this argument presumes that the nonequilibrium steady state is fluid, with currents $V$ on the order of the bare drift velocity $V_0$. It therefore does not apply at conditions where jamming occurs, e.g. at large $\rho$. There, $D_{\rm eff}$ increases less rapidly than linearly with \pp number; see \f{figSI4}. To address this case one could return to \eq{eq3} and consider $V$ and $D$ to have a nontrivial dependence upon Pe. (As an aside, we note that if in \eq{eq3} we assume $D$ to depend linearly on Pe, which the data of \f{figSI4} suggest is true for some range of Pe, then $D_{\rm eff}$ is linear in Pe.)

Previous work has shown that a microscopic analysis of the oppositely-driven particle system implies laning via a dynamic instability\c{chakrabarti2003dynamical,chakrabarti2004reentrance} or the development of anisotropic particle correlations in the disordered phase\c{kohl2012microscopic}. Our approximate argument complements those approaches, suggesting a general and detail-insensitive origin for lane formation. It also motivates the analysis of the following section, in which we discuss the macroscopic consequences of lane formation.

\section{A possible lattice-based reference system for lane formation}
\label{ising}

{\bf Particle drift induces effective interparticle interactions.} Laning occurs because the diffusion constant of a particle can be larger when surrounded by particles of the opposite color than when surrounded by particles of the same color. In \s{results} we argued that this enhancement of diffusion results from the geometric constraint that oppositely-moving particles must, in the time taken to drift into contact, diffuse laterally by about a particle diameter. Supporting this argument, the scaling of diffusion rate with \pp number in quasi-single-particle simulations is consistent with our numerics across a broad range of parameters (\f{fig4} and \f{figSI4}). The lattice-based model (Protocol III) shows no tendency to lane because particle motion on scales less that a particle diameter is not represented, and so no enhancement of particle diffusion constant can occur. However, we argue in this section that there {\em does} exist a lattice-based system that one could use as a reference for the off-lattice model, so clarifying the macroscopic behavior of the latter.

The starting point for this analogy is the observation that hard particles with environment-dependent diffusion rates resemble interacting particles. Consider \f{fig_ising1}, which indicates the movement of a shaded particle between two positions, labeled `initial' and `final'. Suppose that particles in this picture possess only hard-core repulsions, and that particles hop uniformly to any location within a specified range of their starting position. Let this rate of hopping be proportional to a function $f$ of the environment of the particle prior to its hop, provided that the hop causes no hard-core overlaps. The ratio of rates at which the shaded particle moves between its initial (i) and final (f) positions is $f_{\rm i}/f_{\rm f}$. For hard particles the ratio of Boltzmann weights between initial and final microstates is unity, i.e. hopping rates do not satisfy detailed balance with respect to the energy function of the system. However, we can consider that hopping rates satisfy detailed balance with respect to {\em some} energy function ${\cal H}$, i.e. we are free to write
\beq
\label{ratio}
\frac{f_{\rm i}}{f_{\rm f}} = \exp\left(-\beta [{\cal H}_{\rm f}- {\cal H}_{\rm i}]\right).
\eeq
In other words, ${\cal H}$ is the particle-particle interaction potential that {\em would} -- in thermal equilibrium and for particles that possess hopping rates insensitive to their environment -- effect the ratio of hopping rates specified on the left-hand side of \eqq{ratio}. Therefore, hard particles with environment-dependent hopping rates $f$ are equivalent to hard particles with environment-independent hopping rates {\em and} interactions of strength
\beq
\label{map}
{\cal H} = \kt \ln f
\eeq
in thermal equilibrium.

A recent paper by Sear\c{sear2015out} demonstrated this equivalence for a lattice model with diffusion rates $f(n) = {\rm e}^{-\alpha n}$, $n$ being the number of nearest neighbors of a given particle. The equivalent equilibrium system is the Ising lattice gas with coupling constant $\alpha$.
\begin{figure}[]
	\includegraphics[width=0.6\linewidth]{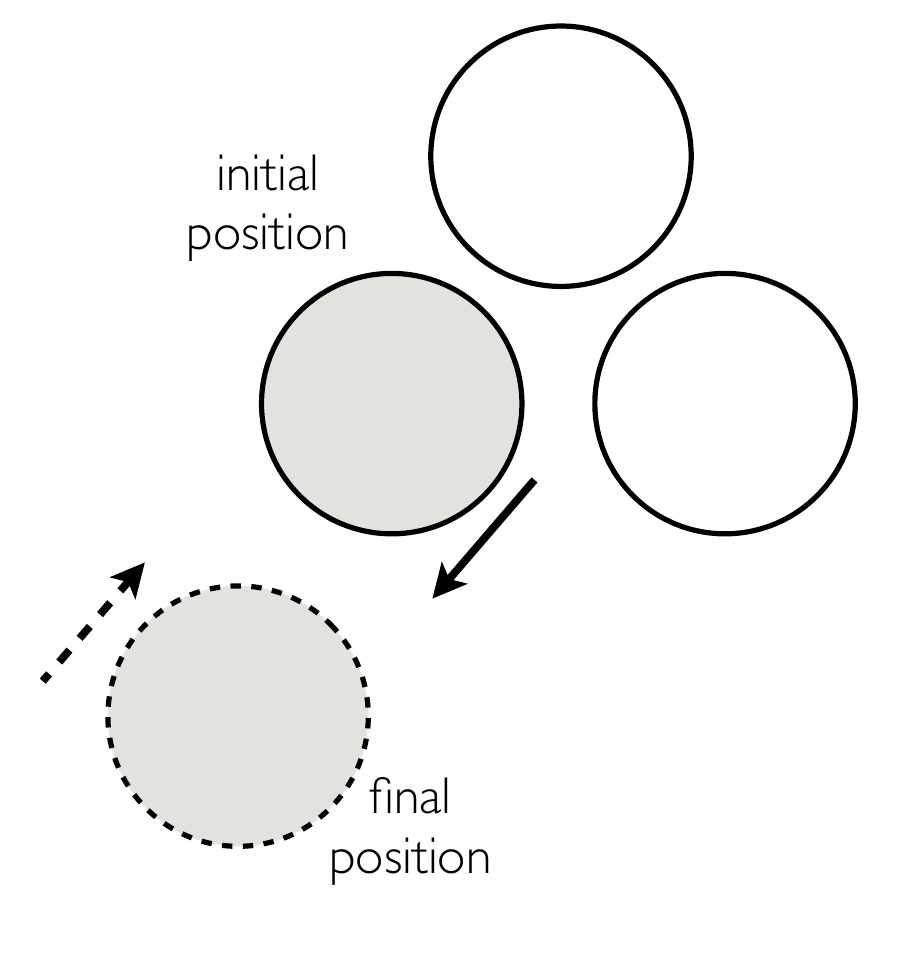}
	\caption{\label{fig_ising1} Diagram used to demonstrate the statistical equivalence between hard particles with environment-dependent hopping rates $f$ and hard particles with environment-independent hopping rates and interactions of strength $\kt \ln f$.}
\end{figure}

\begin{figure*}[]
	\includegraphics[width=.9\linewidth]{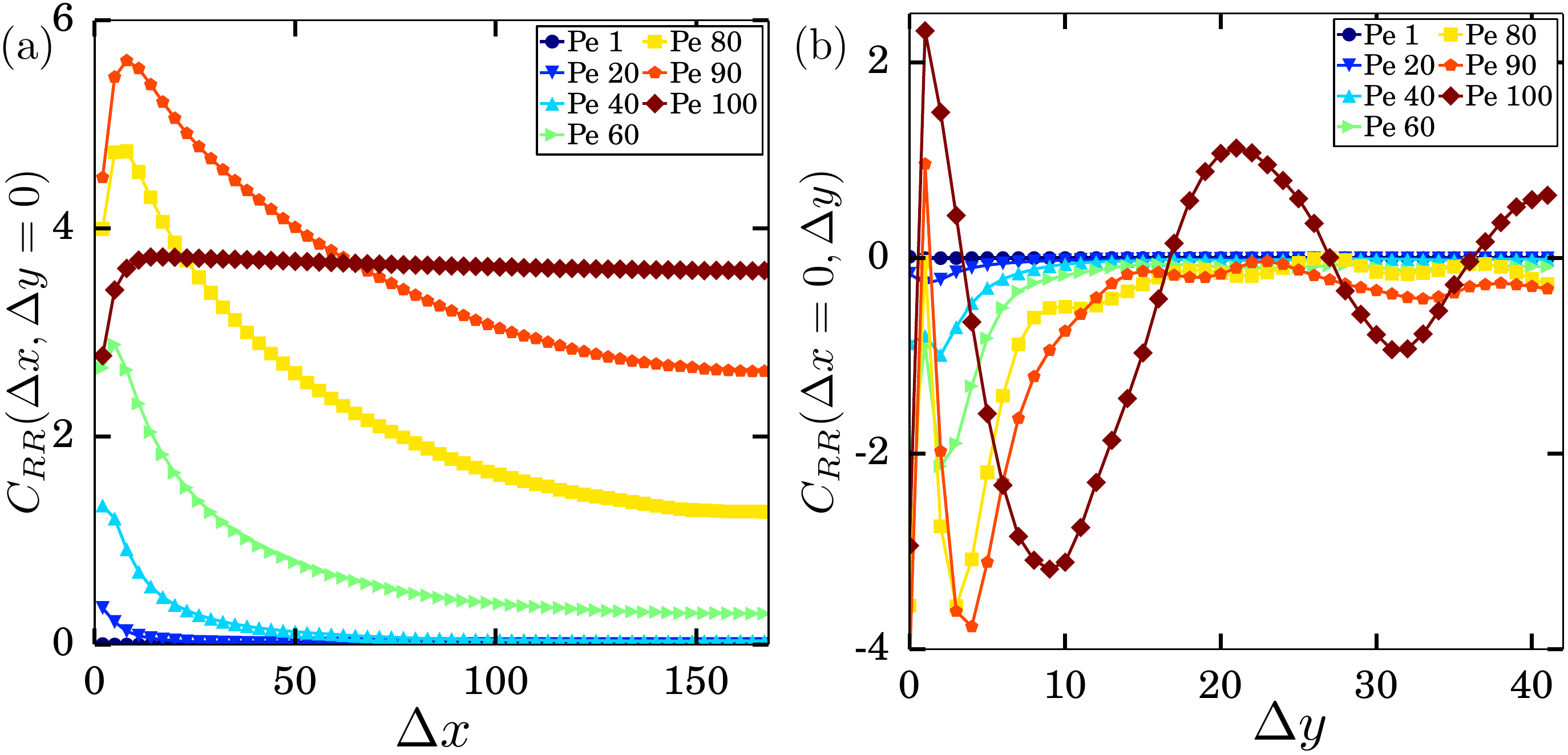}
	\caption{\label{fig_lengths2} Velocity correlation functions, \eqq{velocity_correlation}, along (a) and against (b) the direction of the drive, measured in a simulation box of size $336 \sigma \times 84 \sigma$ at $\rho=0.5$. Long-range correlations in panel (a) are evident well below the critical \pp number of 90. Above this value the function plotted in (b) acquires an oscillatory structure, signaling the formation of lanes parallel to the driven direction.}
\end{figure*}

The connection made by \eqq{map} has significance for the present problem because the opposing drift of particle types generates environment-dependent diffusion rates (in addition to causing persistent motion): blue particles diffuse more rapidly when near red particles than when not near red particles. One can therefore consider the opposing drift of opposite particle types to generate an effective red-blue repulsion, because blue particles have a tendency to spend more time in the vicinity of blue particles then in the vicinity of red particles. This repulsion must be strongly anisotropic, because only particles in danger of colliding head-on must diffuse unusually rapidly~\footnote{Another potential source of anisotropy is a difference between the basic rate of longitudinal diffusion (in the direction of the field) and the basic rate of lateral diffusion (against the field). Our numerics show these quantities to increase with \pp number in a similar but not identical fashion.}. Given the emergence of an effective interparticle interaction and the presence of persistent particle motion, it is natural to consider the simplest model that possesses both features, the driven Ising lattice gas (DLG)\c{katz1984nonequilibrium}. In this model Ising spins move under the influence of an `electric field' $E $ that drives spin types (or particles and holes in lattice-gas language) in opposing directions. The half-full DLG displays a continuous order-disorder phase transition, with non-Ising exponents, at a critical temperature that increases with $E$ and saturates as $E \to \infty$ at about 1.4 times the Ising critical temperature\c{schmittmann1990critical,saracco2003critical,zia2010twenty}. 

It is likely that the off-lattice model resembles the DLG most closely under incompressible conditions, i.e. when the off-lattice model does not exhibit large density fluctuations. Our simulations indicate that while the off-lattice model does exhibit large density fluctuations in certain parameter regime, lane formation can be seen under approximately incompressible conditions. Under such conditions a red-blue repulsion is equivalent to red-red and blue-blue attractions that are more favorable than the red-blue interaction, similar to the ferromagnetic Ising model interaction hierarchy. We then suggest that an appropriate DLG representation of the off-lattice model is one in which the electric field $E \propto $ Pe; the Ising magnetic field is zero (appropriate to red-blue equimolar conditions); and the horizontal $J$ (driven-direction) and vertical $J'$ (lateral) Ising couplings are unequal, and scale approximately logarithmically with \pp number (see \s{onsager}).

This analogy suggests that the emergent behavior of the off-lattice model of lane formation should be similar, as \pp number is increased, to that of the DLG as temperature is decreased and electric field increased. Consistent with this suggestion we found the following qualitative similarities between the two models.\\

{\bf The off-lattice model exhibits long-range correlations in the homogeneous phase.} The DLG exhibits long-range correlations in the homogeneous phase: structural two-point correlations decay as $r^{-2}$ in two dimensions\c{garrido1990long,schmittmann2000viability}. We note that structural two-point correlations in the off-lattice driven model show power-law decay consistent with $r^{-2}$ scaling\c{kohl2012microscopic}. To demonstrate that dynamic quantities also show long-range behavior in the homogeneous phase we applied to the off-lattice driven model an order parameter designed to measure velocity correlations between particles separated by the vector $(\Delta x, \Delta y)$,
\begin{widetext}
\beq
\label{velocity_correlation}
C_{\rm RR}(\Delta x, \Delta y) \equiv \langle \frac{1}{ \mathcal{N}}\sum_{i=1}^{N_{\rm R}-1}\sum_{j=i+1}^{N_{\rm R}} v_i(x_i,y_i)v_j(x_j,y_j) \delta(|x_i-x_j|,\Delta x)\delta(|y_i-y_j|,\Delta y)\rangle - \langle \frac{1}{N_{\rm R}}\sum_{i=1}^{N_{\rm R}} v_i(x_i,y_i)  \rangle^2.
\eeq
\end{widetext}
Here $\mathcal{N}$ is the normalization
 \beq
 \mathcal{N}\equiv\sum_{i=1}^{N_{\rm R}-1}\sum_{j=i+1}^{N_{\rm R}} \delta(|x_i-x_j|,\Delta x)\delta(|y_i-y_j|,\Delta y).
 \eeq
In \eqq{velocity_correlation} the subscript RR indicates correlations between red particles (by symmetry, the blue-blue correlation function shows similar behavior); $v_i$ is the coarse-grained velocity of (red) particle $i$ over time $\tau$ (time over which a particle at low \pp number in vacuum will drift on the order of $\sigma$); the sums run over red particles ($N_{\rm R}$ is the total number of red particles); $\delta$ is the Dirac delta; and averages $\langle \cdot \rangle$ are taken over dynamical trajectories. In \f{fig_lengths2} we show that velocity correlation functions in driven- and non-driven directions reveal the emergence of correlations that are of substantial range, of order that of the simulation box, for values of \pp number below the critical value (note that the critical value of Pe varies with simulation box size and shape). Velocity correlations that oscillate in the non-driven direction reflect the incipience of persistent lanes that become stable above critical driving.
 
 In the ordered phase we estimate that the drive-induced effective interparticle interactions {\em alone} imply the emergence of structures whose sizes grow algebraically with \pp number (in a finite simulation box); see \s{onsager}. This estimate is rough, because this scaling is presumably modified by the presence of persistent particle motion, but in a way that is currently not known.\\

{\bf The off-lattice driven model exhibits system-spanning fluctuations and a change of slope of particle current with \pp number.} The half-full DLG displays a continuous phase transition characterized by system-spanning fluctuations and a discontinuity in the rate of change of particle current with temperature\c{marro1996,zia2010twenty}. By analogy, we expect the off-lattice driven system to show a regime of system-spanning fluctuations as Pe is made large, and a change of slope of particle current with \pp number. In \f{fig_fluctuations} we show that both features are seen. Current is defined per-particle as $\Delta x_i(\tau)\equiv x_i(t+\tau)-x_i(t)$, where $\tau$ is a coarse-graining time over which a particle at low \pp number in vacuum will drift on the order of $\sigma$.\\

\begin{figure}[]
	\includegraphics[width=\linewidth]{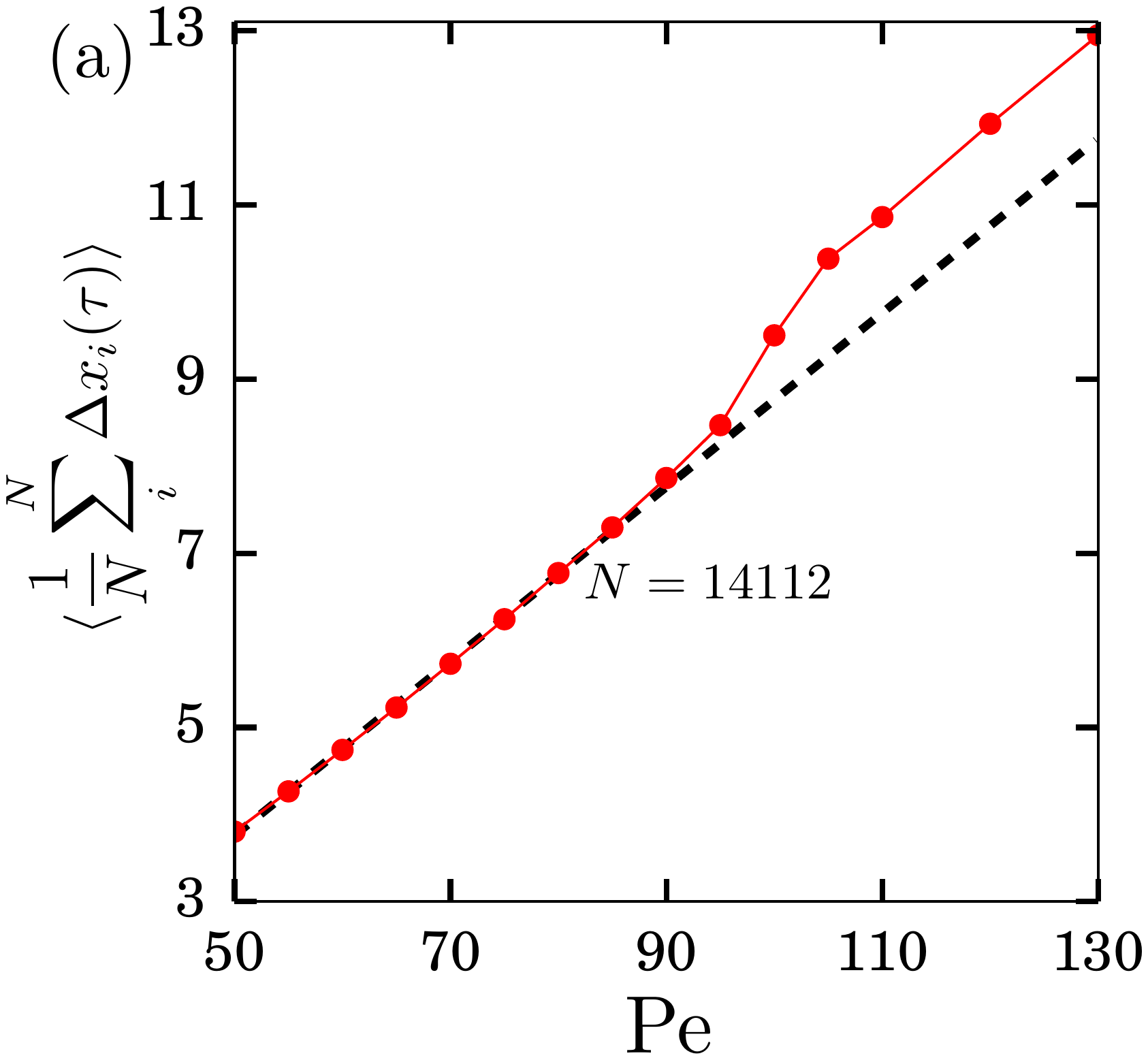}
	\includegraphics[width=\linewidth]{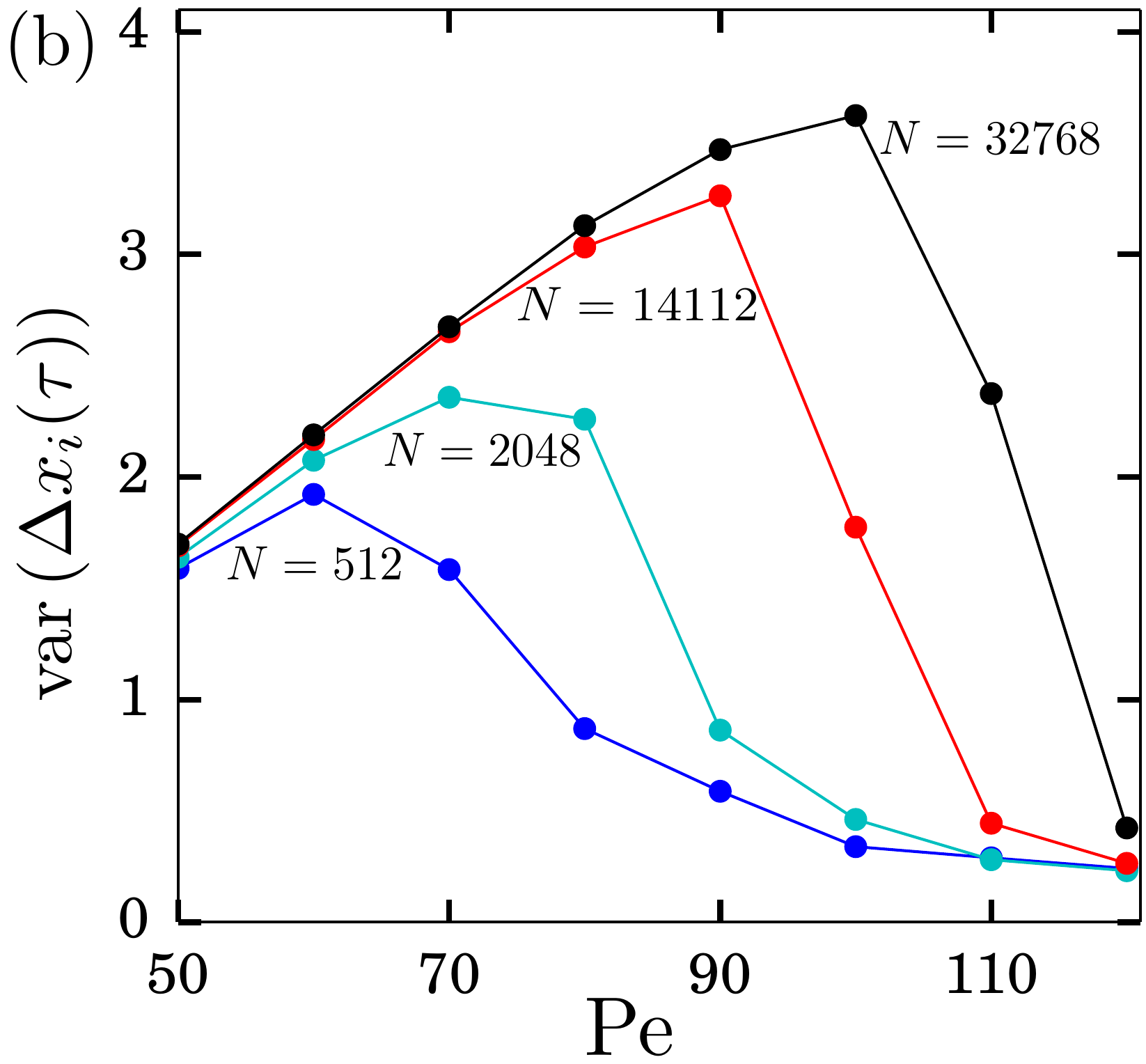}
	\caption{\label{fig_fluctuations} The off-lattice model displays (a) a change of slope of particle current with Pe (the dashed black line shows a linear fit to the current values for \pp numbers 50-85 in order to highlight the change in slope) and (b) system-spanning current fluctuations, similar to the behavior of the driven lattice gas at its critical point. $N$ indicates the number of particles present in the simulation box. Fluctuations of individual particle diffusion constants behave similarly (see \f{figSI3}).}
\end{figure}
\begin{figure*}[]
	\includegraphics[width=0.75\linewidth]{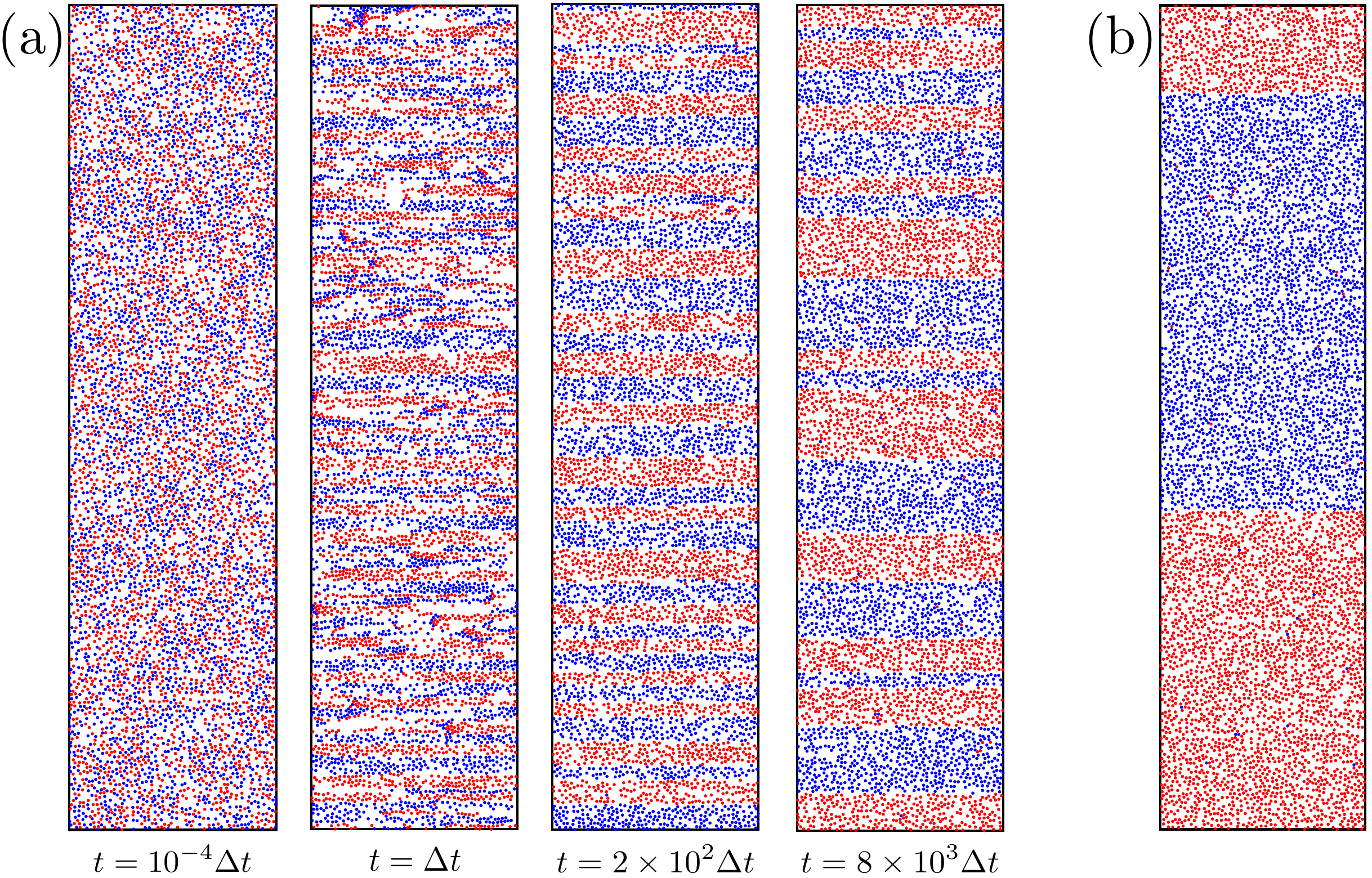}
	\caption{\label{fig_pictures}The off-lattice model displays, in the two-phase region, a slow coarsening process similar to that seen in the driven lattice gas. Panel (a) shows a coarsening process starting from disordered initial conditions ($\Delta t$ is the Langevin timestep). Panel (b) shows a snapshot obtained by choosing as the initial condition two lanes, which persisted upon simulation.}
\end{figure*}

{\bf Macroscopic consequences of lane formation.} The emergence of an effective interparticle attraction and the DLG analogy strongly suggest the potential for macroscopic phase separation in the off-lattice driven system. However, under conditions for which macroscopic phase separation is viable, the time to establish phase separation in the DLG diverges with system size\c{levine2001ordering}. By analogy we conjecture that macroscopic phase separation in the off-lattice driven model is in principle viable, meaning that macroscopic domains would persist once formed, but would not be seen in finite time upon starting from disordered initial conditions. The latter conclusion is consistent with the conclusion of Ref.\c{glanz2012nature}, that lane formation begun from disordered conditions does not look like a phase transition. 

To support our conjecture we show in \f{fig_pictures} time-ordered snapshots of the off-lattice driven model above the critical \pp number. Two lanes persist if built `by hand', but do not emerge on the timescale of simulations that are begun from disordered initial conditions. The slow coarsening process seen in our simulations is qualitatively similar to that seen in the DLG\c{levine2001ordering} (see Fig. 2 of\c{hurtado2003kinetics}), and so we expect it to proceed to completion over a time $\sim L_x L_y^3$\c{levine2001ordering}, where $L_x$ is the driven direction. To see this, note that the characteristic timescale for one stage of coarsening, two bands of width $\ell_y$ merging, is $t\sim \ell_y^3 L_x$ (see Ref.\c{levine2001ordering}). The coarsening time is dominated by the last stage, when $\ell_y$ is on the order of $L_y$, which gives a total time $\sim L_x L_y^3$. 

Other authors have noted macroscopic features held in common between the DLG and off-lattice driven models. In particular, interfaces between phases in the DLG can be statistically flat\c{zia2010twenty,leung1993anomalous} even in two dimensions, unlike interfaces in the Ising model which are rough\c{abraham1976interface}. Similarly flat interfaces have been observed\c{interfaces2016} in an off-lattice model of driven particles that shares some basic ingredients with the model studied here.

\section{Conclusions}
\label{conc}

We have studied lane formation in a system of oppositely-driven model colloidal particles using a combination of simulation methods and approximate physical analogies. We argue that the microscopic origin of laning, several aspects of which have been determined previously\c{chakrabarti2003dynamical,chakrabarti2004reentrance,kohl2012microscopic}, can be understood from a simple geometric argument that implies an environment-dependent particle mobility scaling linearly with \pp number. Given that one can equate environment-dependent mobilities with an effective interparticle attraction, we conjecture that the basic features of pattern formation in the off-lattice driven system should be similar to those of the driven lattice gas, whose coupling constants grow approximately logarithmically with \pp number. Consistent with this conjecture we see in simulations of the off-lattice driven model long-rage correlations in the homogenous phase; critical-like fluctuations and a change of slope of particle current with \pp number; and phase separation at large \pp number that persists once formed but takes a long time to develop from disordered initial conditions. There are likely to be important differences between the DLG and the off-lattice driven model, particularly where the latter exhibits large density fluctuations or jamming, but there also exist clear similarities between the two models. It will be valuable to determine the extent to which the DLG can be used as a reference model for other driven systems. Note that lane formation is also seen in 3-dimensional systems \cite{dzubiella2002lane} and it would be interesting to look for evidence of DLG-like behavior there. In addition, the identification that laning results from rectification of diffusion suggests an intriguing connection between the emergent phenomena of driven molecular systems and those of social dynamics, which have been described in similar geometric terms\c{oliveira2016keep}.

\acknowledgements
We acknowledge valuable discussions with Todd Gingrich, Dibyendu Mandal, Suriyanarayanan Vaikuntanathan, Robert L. Jack, C. Patrick Royall, John Edison, Thomas Speck, and Grzegorz Szamel. KK acknowledges support from the NSF Graduate Research Fellowship. PLG was supported by the U.S. Department of Energy, Office of Basic Energy Sciences, through the Chemical Sciences Division (CSD) of the Lawrence Berkeley National Laboratory (LBNL), under Contract DE-AC02-05CH11231. This work was done as part of a User project at the Molecular Foundry at Lawrence Berkeley National Laboratory, supported by the Office of Science, Office of Basic Energy Sciences, of the U.S. Department of Energy under Contract No. DE-AC02--05CH11231.


\appendix
\onecolumngrid

\renewcommand{\theequation}{S\arabic{equation}}
\renewcommand{\thefigure}{S\arabic{figure}}
\renewcommand{\thesection}{S\arabic{section}}

\setcounter{equation}{0}
\setcounter{section}{0}
\setlength{\parindent}{0pt}%

\section{Simulation Details}

\subsection{Biased Off Lattice Monte Carlo Simulations}
\label{monte}

Protocol II described in the main text is a Metropolis Monte Carlo simulation in which particle displacements are drawn uniformly from within a square of side $2L$ centered at $(\pm c,0)$ (the upper and lower sign applying for blue and red particles, respectively). For an isolated red particle we then have, for unit time, 
\beq
\av{x}= \frac{1}{2L} \int_{-L-c}^{L-c} x \, {\rm d}x = -c.
\eeq
Thus $v_x = -w_0 c$, where $w_0$ is a basic rate. In the perpendicular direction we have $\av{y}=0$ and $v_y=0$. Thus $v \equiv \sqrt{v_x^2+v_y^2}=w_0 c.$

The mean-squared displacement of an undriven particle (or of a driven particle in its rest frame) in either direction in unit time is
\beq
\av{x^2}_0 = \frac{1}{2L} \int_{-L-c}^{L-c} x^2 \, {\rm d}x = \frac{L^2}{3},
\eeq
giving a bare diffusion constant $D_0 = w_0 L^2/6$.

We define \pp number 
\beq
{\rm Pe} \equiv \frac{v \sigma}{D_0} = \frac{6 c \sigma}{L^2},
\eeq
where $\sigma$ is the particle diameter.

\subsection{Biased Lattice Simulations} Our lattice simulations consist of hard particles (equal in size to the lattice site) with volume exclusion. Monte Carlo moves are local hops with probability $\gamma$ in the $\pm y$ (non-driven) directions, probability $\gamma + \Delta$ in the $+(-)x$ direction for blue (red) particles, and probability $1-(3\gamma+\Delta)$ in the $-(+)x$ direction for blue (red) particles. A hop attempt is rejected if the chosen site is already occupied. No particle swap moves are allowed. These dynamics preserve the number of red and blue particles in the simulation box. We constrained the bare diffusion constants in the $x$ and $y$ directions to be the same. For these simulations the \pp number is ${\rm Pe}=\sqrt{1-4\gamma}\sigma/\gamma$. We confirmed that measurements of $\langle x \rangle \sigma / \langle \delta x^2 \rangle$ gave us the expected Peclet number for an isolated particle.

\subsection{Steady State Regimes}
The activity $A(\tau)$ and $\phi$ were used to characterize the steady states. Off-lattice MC simulations were run at a range of step sizes. Structures were labeled `jammed' when $A(\tau)/A(\tau)_{\text{bare}} < 0.5$ where $A(\tau)_{\text{bare}}$ is the activity of an isolated particle. For Langevin simulations and Monte Carlo simulations with step sizes larger than $0.005 \sigma$, structures were labeled `laned' when $\phi$ was larger than 0.5. For step sizes smaller than this, simulations equilibrated extremely slowly and often did not reach a stable value of $\phi$ over the course of $5 \times10^{10}$ Monte Carlo sweeps. To approximate the boundary between laned and disordered states for these step sizes (the dashed red line in \f{fig1}), we used the criterion that $\phi$ (without time-averaging) reach a value of 0.4 or larger at some point during the trajectory.  

We found that Monte Carlo simulations of WCA particles showed similar behavior to Langevin dynamics (laning above \pp 80 and no jamming) at step sizes $\Delta _x/ \sigma < 0.01$. Hard disks required a smaller step size, $\Delta _x/ \sigma < 0.005$, to show behavior similar to Langevin dynamics. \f{fig1} shows the steady-state regimes for hard disks; the diagram would look similar for WCA particles, but with the jammed/flowing boundary shifted to a higher step sizes.

\section{Thermodynamic Perturbation Theory, WCA particles to hard disks} In figure 2 of the main text we compare the results of Brownian dynamics simulations of soft (WCA) particles and Monte Carlo simulations of hard disks. We chose a hard disk radius such that the thermodynamics of the two systems are equivalent (in the sense described below). We verified that little difference is seen in MC simulations upon small variations of disk diameter.

The free energy of a collection of interacting particles is a functional of the pair potential:
\begin{equation}
A[u(r)]=-k_BT\ln \int dr^N e^{ -\beta \frac{1}{2}\sum_{i \neq j}u(r_{i,j})}=-k_BT\ln \int dr^N \prod_{i,j}f(r_{i,j}),
\end{equation}
where $f(r)$ is the Meyer f-function
\begin{equation}
f(r)=e^{-\beta u(r)}.
\end{equation}

Referring to the WCA pair potential with the subscript ${\rm o}$ and the hard disk pair potential with the subscript ${\rm d}$, we want to make their free-energy functionals as close as possible, i.e.

\begin{equation}
A_{\rm o}=A_{\rm d}+\Delta A
\end{equation}

with $\rm d$ chosen such that $\Delta A \approx 0 $.

We can define

\begin{equation}
f_{\lambda}(r) = f_{\rm d}(r)+\lambda\Delta f(r)
\end{equation}

with $\Delta f = f_{\rm o}(r)-f_{\rm d}(r)$. Then

\begin{equation}
\Delta A = A(\lambda =1)-A(\lambda =0) = \int_0^1 d\lambda \int d \vec{r}\frac{\delta A}{\delta f_{\lambda}(r)\Delta f(r)}
\end{equation}

where

\begin{equation}
\frac{\delta A}{\delta f_{\lambda}(r)\Delta f(r)}=-k_BT \frac{1}{Q}\int d r^N \frac{1}{2}N(N-1)\left[ \prod_{i,j \neq (1,2)}f(r_{i,j}) \right] \delta (r-r_{1,2}).
\end{equation}

Choose particles $i$ and $j$ as 1 and 2

\bea
&=&-\frac{1}{2}k_BT N^2 \frac{1}{Q}\int dr^N \left[ \prod_{i,j}f(r_{i,j}) \right]e^{\beta u(r_{1,2})}\delta(r-r_{12}) \nonumber \\
&=&-\frac{1}{2}k_BTN^2 \frac{ \langle \delta(r-r_2)\delta(r_1) \rangle}{\langle \delta(r_1) \rangle}. 
\eea

Using

\begin{equation}
\frac{ \langle \delta(r-r_2)\delta(r_1) \rangle}{\langle \delta(r_1) \rangle} = \frac{g(r)\bar{\rho}}{N}
\end{equation}

where $\bar{\rho}$ is the average density of the system leaves us with

\begin{equation}
\frac{\delta A}{\delta f_{\lambda}(r)\Delta f(r)}=-\frac{1}{2}k_BTN\bar{\rho}e^{\beta u(r)}g(r).
\end{equation}

Note that $e^{\beta u(r)}g(r)=y(r)$, the cavity distribution function. This gives:

\begin{equation}
\Delta A = \int_0^1 d \lambda \int d \vec{r} \left[ -\frac{1}{2}k_BTN\bar{\rho} \right] y_{\lambda}(r)\delta f(r)
\end{equation}

which we want to set to 0. If $\rm d$ is chosen well, $y_{\lambda}(r) \approx y_{\rm d}(r)$, so we only need to worry about

\begin{equation}
\int d \vec{r} y_{\rm d}(r)\Delta f(r) =0.
\end{equation}

This brings us to Perkis-Yevick Theory:

\begin{equation}
h(r)=c(r)+\bar{\rho}\int dr' c(r-r')h(r')
\end{equation}

where $h(r)=g(r)-1$ and $c(r)$ is the direct correlation function. As shown in\c{hansen1990theory}

\begin{equation}
y(r)-1 \approx \bar{\rho} \int d r' c(r-r')h(r') = h(r)-c(r).
\end{equation}

For $r< \rm d, h(r)=-1$ so $y(r)=-c(r)$ leaving us with

\begin{equation}
\int d\vec{r}y_{\rm d}(r)\Delta f(r)=0
\end{equation}

where $\Delta f(r) = e^{-\beta u_{\rm o}(r)}-\theta(r-\rm d)$.

Percus-Yevick theory predicts a form for $c(r)$ that has been solved analytically in 3 dimensions, but to the best of our knowledge not in 2 dimensions, so we numerically calculated $y_{\rm d}(r)$. It turns out that a hard disk diameter of $\sigma$ is a good approximation for WCA particles of diameter $\sigma$, at least when comparing the free energy functionals.

\section{Off-lattice model-DLG analogy, and approximate lengthscales in the ordered phase}
\label{onsager}

The analogy drawn in the main text suggests that the off-lattice model can be related to the DLG whose Ising couplings scale roughly as
\beq
\label{horiz}
2 J \propto \kt \ln  (1+ {\rm Pe})
\eeq
and
\beq
\label{vert}
2 J' \propto \kt \ln (1+\lambda {\rm Pe}),
\eeq
for bonds running in driven and non-driven directions, respectively. Here $\lambda<1$ is a geometric parameter that could be fixed by requiring the model to be critical at a particular value of Pe. At the level of the Ising model we can follow Onsager's analysis\c{onsager1944crystal} to show that such couplings imply in the ordered phase the emergence of structures whose characteristic lengthscales grow algebraically with Pe. Assume that the simulation box dimensions are $L_x$ and $L_y$ in driven and non-driven directions. The Ising model surface tension in driven and non-driven directions is $\sigma' = 2J' + \kt \ln \tanh(\beta J)$ and $\sigma = 2J + \kt \ln \tanh(\beta J')$. The free-energy cost required to create a vertical boundary of length $L_y$ is $\sigma L_y$, and so the characteristic length $l_x$ between such boundaries is the exponential of this quantity multiplied by $\beta$, i.e. $l_x = \left({\rm e}^{2 \beta J} \tanh \left( \beta J'\right) \right)^{L_y}$ (this result is Equation (124) of Ref.\c{onsager1944crystal}; note that the version of this result quoted in the abstract of that paper appears to have a spurious factor of 2 within the $\tanh$ function). Inserting into this expression the couplings \eq{horiz} and \eq{vert}, with constants of proportionality taken to be unity, we find the characteristic domain length in the driven direction to be
\beq
\label{onsager_horiz}
l_x =\left((1+{\rm Pe})\frac{\lambda {\rm Pe}}{\lambda {\rm Pe}+2}\right)^{L_y}.
\eeq
For large Pe this length grows as a power law, $l_x \sim {\rm Pe}^{L_y}$ (taking non-unit constants of proportionality in \eq{horiz} and \eq{vert} modifies the exponent, but does not change the fact that the lengthscale goes as a power of \pp number).

The characteristic length of domains in the non-driven direction, i.e. the equilibrium lane width, can be found in similar fashion; it is   
\beq
\label{onsager_vert}
l_y =\left((1+\lambda {\rm Pe}) \frac{{\rm Pe}}{{\rm Pe}+2}\right)^{L_x},
\eeq
which for large Pe grows as $l_y \sim (\lambda {\rm Pe})^{L_x}$.

These results are consistent in a general sense with the results of Ref.\c{glanz2012nature}, whose authors measured a lengthscale within the off-lattice driven model that for large Pe grows with Pe either exponentially or as a power law. However, the connection is not a precise one because that lengthscale is neither of the Onsager lengths stated here. In addition, the above analysis concerns the {\em undriven} Ising model, and the driven version (the DLG) possesses anisotropy of domains on account of the drive, even for identical couplings $J = J'$\c{schmittmann1990critical}. Interfaces in the DLG are also statistically smoother than those in the Ising model\c{zia2010twenty,abraham1976interface}.

\section{Additional Figures}

The following figures supplement those in the main text, and are called out from there.

\begin{figure}[]
	\includegraphics[width=.75\linewidth]{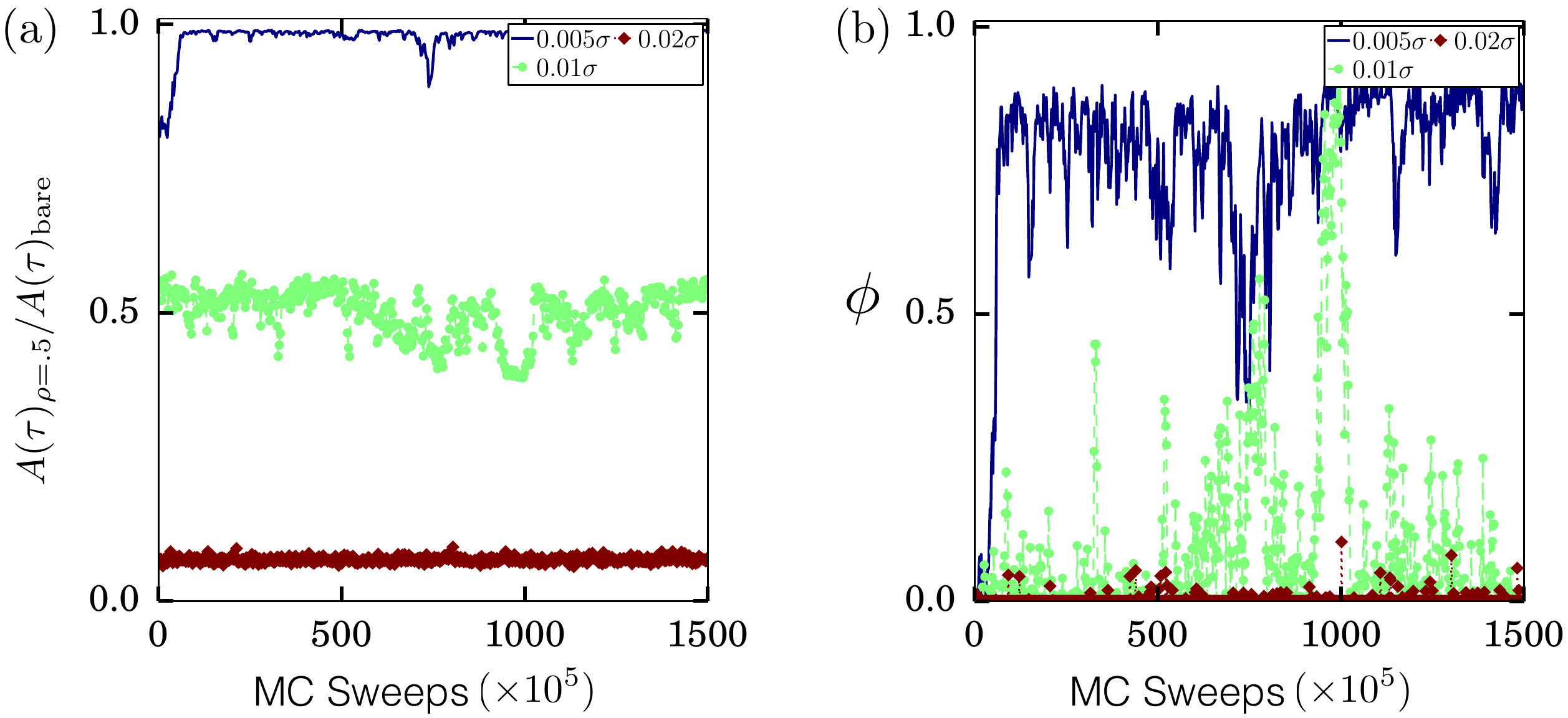}
	\caption{\label{figSI1} Activity (scaled by the average activity for an isolated particle) and the laning order parameter $\phi$ as a function of MC sweep for WCA particles at 3 different step sizes.}
\end{figure}

\begin{figure}[]
	\includegraphics[width=.75\linewidth]{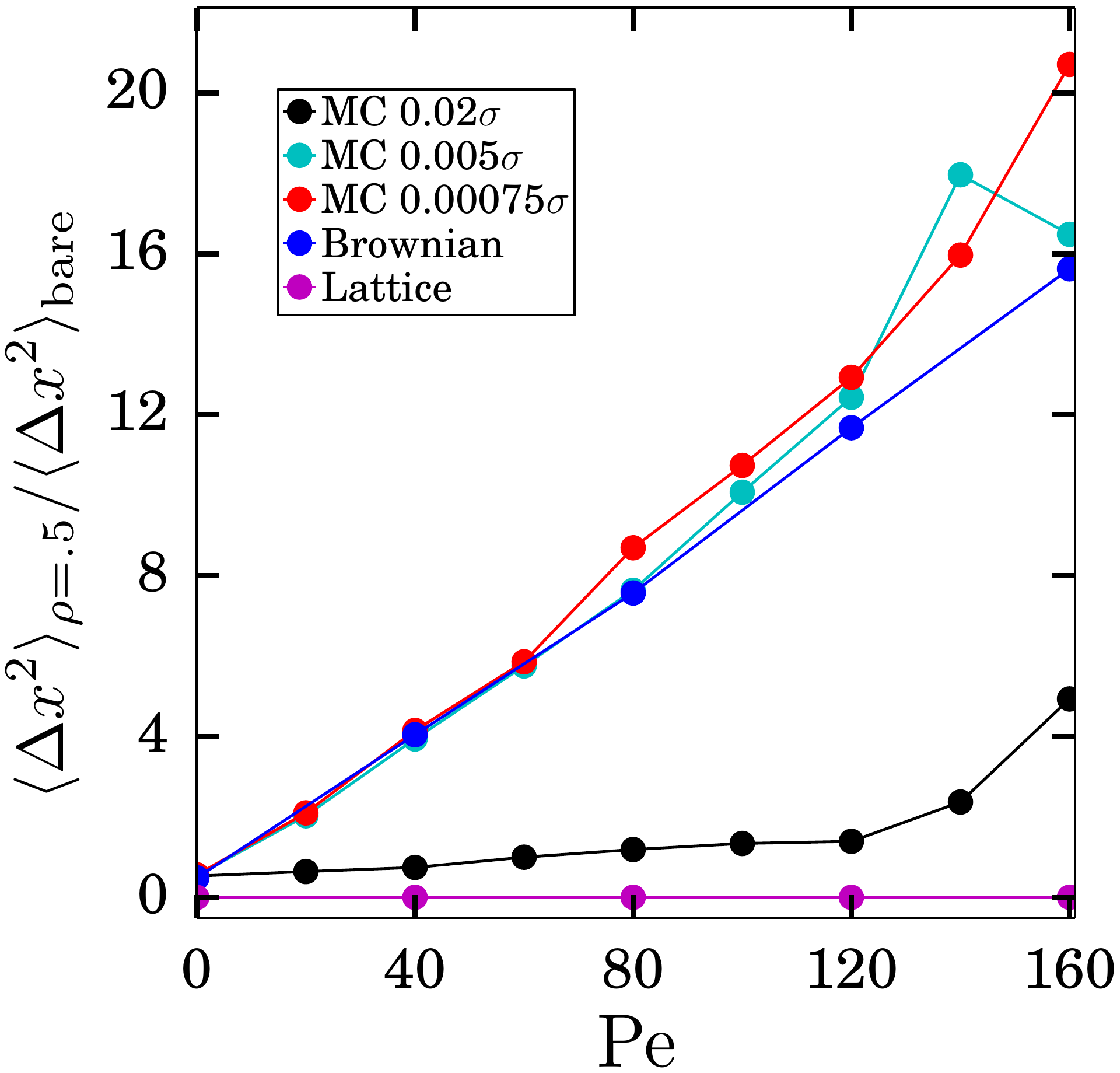}
	\caption{\label{figSI2} The longitudinal mean-squared displacement (in the driven direction) of a test particle in the presence of particles of the other type, normalized by the test particle's bare mean-squared displacement, as a function of \pp number, for different dynamic protocols.}
\end{figure}

\begin{figure}[]
	\includegraphics[width=.75\linewidth]{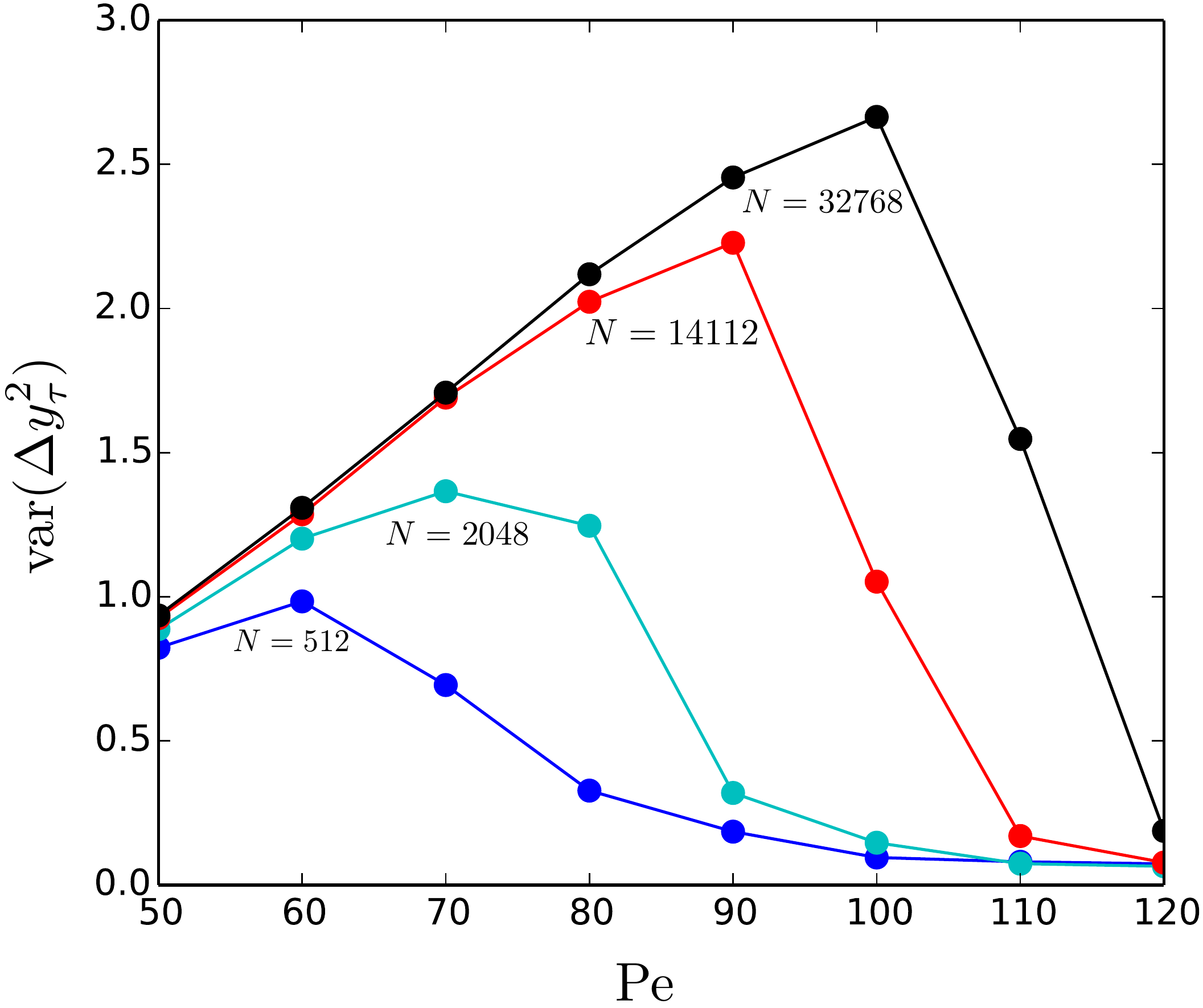}
	\caption{\label{figSI3} Lateral mean-squared displacement (here $\Delta y_{\tau} = y(t)-y(t+\tau)$) distributions broaden near criticality in a manner similar to distributions of particle currents; see \f{fig_fluctuations}.} 
\end{figure}

\begin{figure*}[]
	\includegraphics[width=.85\linewidth]{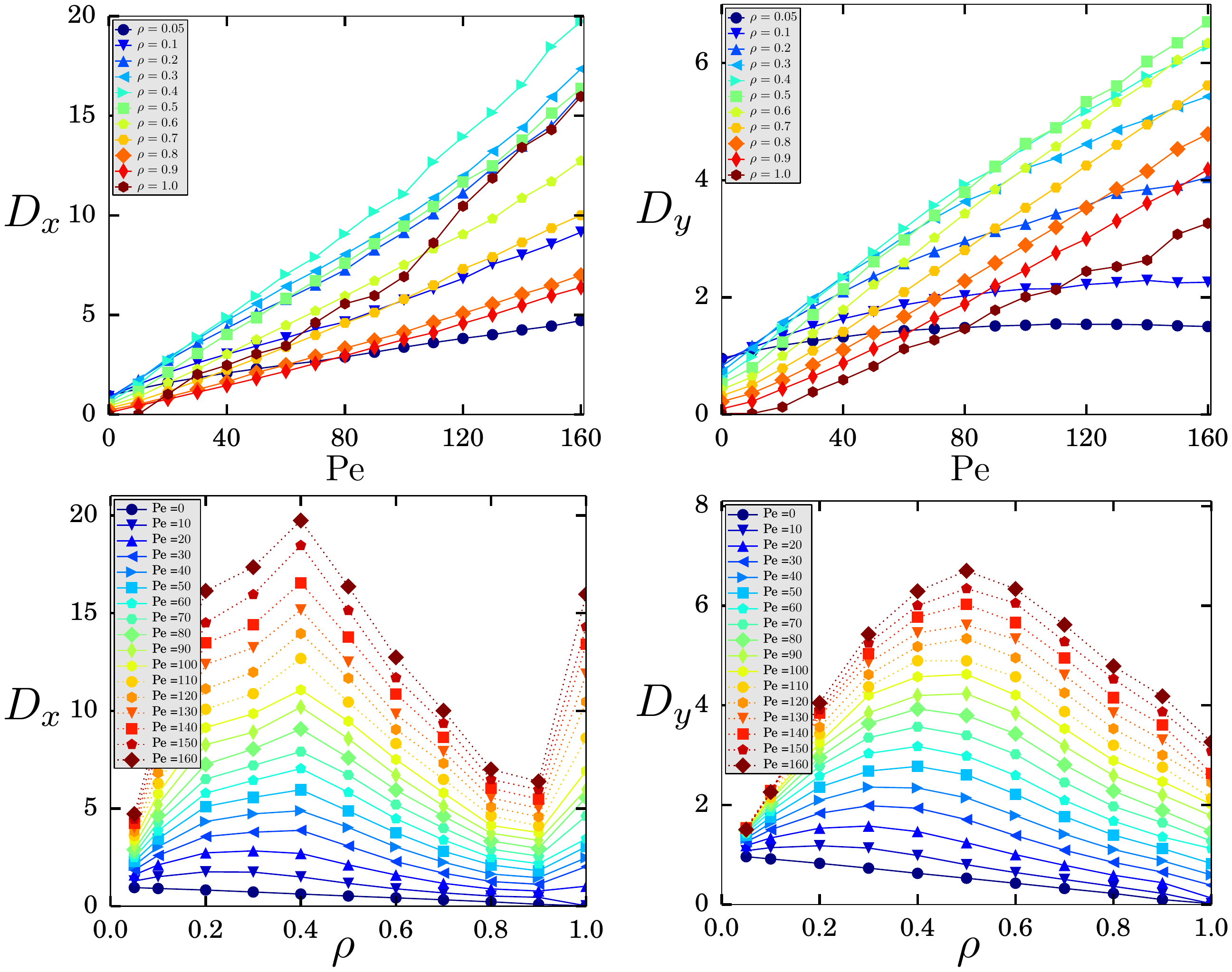}
	\caption{\label{figSI4} Measured diffusion constants for a range of densities and Peclet numbers. The linear scaling of lateral diffusion constant with Pe suggested by \eqq{diff_const} is evident for a range of Pe and $\rho$ (top right panel), and breaks down at large packing fraction and small Pe.}
\end{figure*}

\end{document}